\documentclass[twocolumn,twocolappendix]{aastex631}
\usepackage{float}

\usepackage{hhline}

\begin{document}

\title{JWST+ALMA reveal the build up of stellar mass \\ in the cores of dusty star-forming galaxies at Cosmic Noon}

\correspondingauthor{Sarah Bodansky}
\email{sbodansky@umass.edu}

\author[0000-0001-8246-1676]{Sarah Bodansky}
\affiliation{Department of Astronomy, University of Massachusetts, Amherst, MA 01003, USA}

\author[0000-0001-7160-3632]{Katherine E. Whitaker}
\affiliation{Department of Astronomy, University of Massachusetts, Amherst, MA 01003, USA}
\affiliation{Cosmic Dawn Center (DAWN), Denmark}

\author[0009-0002-5707-2809]{Ayesha Abdullah}
\affiliation{Department of Astronomy, University of Massachusetts, Amherst, MA 01003, USA}

\author[0000-0002-3101-8348]{Jamie Lin}
\affiliation{Department of Physics and Astronomy, Tufts University, Medford, MA 02155, USA}

\author[0000-0001-5851-6649]{Pascal A. Oesch}
\affiliation{Department of Astronomy, University of Geneva, Chemin Pegasi 51, 1290 Versoix, Switzerland}
\affiliation{Cosmic Dawn Center (DAWN), Denmark}
\affiliation{Niels Bohr Institute, University of Copenhagen, Jagtvej 128, K{\o}benhavn N, DK-2200, Denmark}

\author[0000-0001-8592-2706]{Alexandra Pope}
\affiliation{Department of Astronomy, University of Massachusetts, Amherst, MA 01003, USA}

\author[0000-0003-1207-5344]{Mengyuan Xiao}
\affiliation{Department of Astronomy, University of Geneva, Chemin Pegasi 51, 1290 Versoix, Switzerland}

\author[0000-0002-9672-3005]{Alba Covelo-Paz}
\affiliation{Department of Astronomy, University of Geneva, Chemin Pegasi 51, 1290 Versoix, Switzerland}

\author[0000-0002-7031-2865]{Sam Cutler}
\affiliation{Department of Astronomy, University of Massachusetts, Amherst, MA 01003, USA}

\author[0009-0004-1123-001X]{Carlos Garcia Diaz}
\affiliation{Department of Astronomy, University of Massachusetts, Amherst, MA 01003, USA}

\author[0000-0002-2419-3068]{Minju M. Lee}
\affiliation{Department for Space Research and Space Technology (DTU Space), Technical University of Denmark, Elektrovej 327, DK-2800 Kgs. Lyngby, Denmark}
\affiliation{Cosmic Dawn Center (DAWN), Denmark}

\author[0000-0003-0415-0121]{Sinclaire M. Manning} \affiliation{Department of Astronomy, University of Massachusetts, Amherst, MA 01003, USA}

\author[0000-0001-5492-4522]{Romain A. Meyer}
\affiliation{Department of Astronomy, University of Geneva, Chemin Pegasi 51, 1290 Versoix, Switzerland}

\author[0000-0002-7064-4309]{Desika Narayanan}
\affiliation{Department of Astronomy, University of Florida, 211 Bryant Space
Sciences Center, Gainesville, FL 32611, USA}
\affiliation{Cosmic Dawn Center (DAWN), Denmark}

\author[0000-0002-7524-374X]{Erica Nelson}
\affiliation{Department for Astrophysical and Planetary Science, University of Colorado, Boulder, CO 80309, USA}

\author[0000-0003-4702-7561]{Irene Shivaei}
\affiliation{Centro de Astrobiolog\'{i}a (CAB), CSIC-INTA, Carretera de Ajalvir km 4, Torrej\'{o}n de Ardoz, 28850, Madrid, Spain}

\author[0000-0002-8282-9888]{Pieter van Dokkum}
\affiliation{Astronomy Department, Yale University, 52 Hillhouse Ave, New Haven, CT 06511, USA}

\begin{abstract}

Dusty star-forming galaxies have long been suspected to serve as the missing evolutionary bridge between the star-forming and quiescent phases of massive galaxy evolution.  
With the combined power of JWST and ALMA, it is now possible to use high resolution imaging at rest-frame ultraviolet (UV), optical, near-infrared (NIR), and sub-mm wavelengths to study the multi-wavelength morphologies tracing both the stellar populations and dust during this key phase. 
We present the joint analysis of JWST/NIRCam imaging in GOODS-S and mm dust emission traced by ALMA for a sample of 33 
galaxies at $z=1.5$ to $z=5.5$ selected from the 1.1mm GOODS-ALMA 2.0 survey, and compare the morphologies of this population to mass- and redshift-selected samples of field star-forming and quiescent galaxies. The 1.1mm-selected sample is morphologically distinct from other similarly massive star-forming galaxies; we find a steeper size-wavelength gradient from 1.5-4.4$\mu$m, with a more dramatic decrease in size towards longer wavelengths. While the rest-NIR surface brightness profiles of the 1.1mm-selected galaxies are brighter in the inner regions relative to the field star-forming population, they are remarkably similar to the quiescent population. 
These morphological differences could suggest that dusty star-forming galaxies, unlike more typical star-forming galaxies, have already built up stellar mass in a severely dust-obscured core,
leading to extended and clumpy morphologies at rest-UV and rest-optical wavelengths and more compact emission in the rest-NIR that is co-spatial with dust. If the bulge is already established, we speculate that mm-selected galaxies may imminently evolve to join their quiescent descendants.

\noindent 

\end{abstract}

\section{Introduction} \label{sec:intro}
The currently favored scenario to explain the formation and evolution of massive galaxies is one in which star formation and quenching begin in the inner regions of the galaxy and progress outwards. A wide range of observational signatures consistent with this inside-out growth are reported in the literature \citep[e.g.,][]{nelson2016,spilker2019,smith2022, xiao2024, matharu2024}. One physical explanation of this behavior is a model in which disk galaxies acquire gas in their core, catalyzing a rapid phase of star formation and black hole growth \citep[e.g.][]{sanders1988, hopkins2008}. As the central black hole (BH) grows and creates feedback, dust and gas are expelled, triggering an end to star formation. In this paradigm, very dusty star-forming galaxies (DSFGs) are proposed to be the missing link between star-forming and quiescent phases of massive galaxy evolution, and the phase in which bulge formation occurs \citep[e.g.][]{mihos1996,springel2005, gonzalez2011,toft2014, barro2014, dudzeviciute2020,birkin2021,colina2023,long2023}.

Many open questions about the evolutionary status of DSFGs remain. The model discussed by \citet{hopkins2008} suggests that major mergers are key to triggering star formation in DSFGs, which is supported by a range of studies \citep[e.g.][]{tacconi2008, engel2010, hodge2025}. However, recent literature also supports secular processes such as disk instability \citep[e.g.][]{toft2017, gillman2024, faisst2025}, cold accretion \citep[e.g.][]{carilli2010, huang2023}, as well as minor mergers \citep[e.g.][]{gomezguijarro2018} as more likely catalysts of the DSFG phase. Some theoretical models are suggestive of merger-driven star formation \citep[e.g.][]{lower2023}, while others predict a combination of mergers and secular growth \citep{mcalpine2019, lagos2020} or primarily secular growth from accretion \citep{narayanan2015}. Additionally, recent studies find conflicting evidence on whether the growth of stellar mass in DSFGs could occur first in the disk and then in the bulge in an outside-in fashion \citep{lebail2024}, or follows the inside-out model where bulge formation occurs first \citep{kamieneski2023,hodge2025}.  Where and how star formation is induced in the DSFG stage may determine a galaxy’s morphological evolution and quenching timescale \citep{park2022}, so these conflicting results leave the future transformation of DSFGs into quiescent galaxies unresolved.

Understanding stellar mass assembly in DSFGs is complicated in part due to the complex relationship between dust and star formation. Dust and stars are inextricably linked within galaxies. As stars eject matter over the course of their lives, they enrich the ISM with metals that may end up condensing into dust grains \citep{Hirashita2012}. Simultaneously, the births and deaths of stars can devour and destroy dust. Dust also plays a key role in the formation of stars, by enabling a relatively efficient channel for the production of molecular gas \citep{Hollenbach1971}. At the same time, the dusty molecular clouds where stars are born can enshroud younger stellar populations, blocking a direct view of this stellar emission at rest-optical wavelengths \citep{calzetti2001}. Given this complicated relationship, the star-dust geometry in galaxies is variable and difficult to disentangle. Dust is often associated with star-forming regions, and more massive and star-forming galaxies are observed to have larger dust masses \citep{kirkpatrick2017,popping2017, guerrero2023}. Dust can completely obscure our view of stellar populations \citep[e.g.,][]{Huang_2011,fudamoto2021,manning2022,barrufet2023}, and dust emission can be offset from or more compact relative to where stellar emission is observed \citep{simpson2015,hodge2019,gullberg2019,Rujopakarn2023}. These effects are particularly pronounced in the most highly star-forming galaxies, given their vast dust reservoirs \citep{whitaker2017, zimmerman2024}.

The morphologies of both stars and dust together can provide a more complete picture of the assembly of galaxies over cosmic time. Prior to JWST, studies of the galaxy morphologies from stellar light were limited to rest-optical wavelengths beyond \emph{z}$\sim$1 (and only rest-ultraviolet for $z\gtrsim5$). Comparisons between rest-frame optical sizes and dust continuum sizes have shown that dust continuum emission is more compact relative to stellar emission at \emph{z}$\sim$2 \citep{simpson2015,tadaki2017,nelson2019, gullberg2019}. This discrepancy between compact dust emission at the core, which traces regions of active star formation, and extended emission in the rest-optical from existing stellar populations, suggests that galaxies in this epoch may be in the process of bulge formation, with star formation progressing from the inside out prior to quenching. However, the rest-frame optical is heavily impacted by dust attenuation in extreme star-forming galaxies. If dust is preferentially concentrated in a compact core, the rest-frame optical light profiles will flatten at the center and the apparent sizes at these wavelengths consequently become extended due to this dust gradient \citep{chen2022, hodge2025}. Additionally, rest-frame optical morphologies primarily trace short-lived stellar populations at the more massive end of the initial mass function (IMF). Instead, rest-NIR observations are less sensitive to the effects of dust attenuation, and are a better mass-weighted tracer of the bulk stellar emission \citep{clausen2025}. 

With JWST/NIRCam, the rest-NIR can be explored for the first time up to \emph{z}$\sim$5. Working in tandem with ALMA, we can now obtain a resolved multi-wavelength view of this ecosystem: star formation in the rest-optical, stellar mass in rest-NIR, and dust in rest-submm wavelengths. Recent studies find that the rest-NIR sizes of star-forming galaxies at cosmic noon are more compact relative to their rest-optical sizes, but are more extended compared to the dust continuum \citep{chen2022,tadaki2023, gillman2024}, suggesting that stellar mass extends beyond a central region of star formation. JWST/NIRCam has also revealed smooth rest-NIR morphologies that contrast to clumpy emission in the rest-optical, which may either reflect the effects of dust attenuation in the rest-optical or inherently uniform distribution of stellar mass relative to clumps of star-formation \citep{boogaard2024}. Other studies have found evidence that dusty star-forming galaxies are at the onset of bulge formation \citep{lebail2024,hodge2025}.

This recent work has illuminated the rest-NIR morphologies of DSFGs. We can now also study morphologies and bulge formation in the overall massive star-forming galaxy population \citep[e.g.][]{ji2024} compared to quiescent galaxies \citep[e.g.][]{benton2024}. What is lacking is a  direct comparison of the rest-NIR properties of quiescent galaxies, star-forming galaxies, and DSFGs to see whether DSFGs are really an evolutionary bridge in the stellar mass assembly of massive galaxies.

In this paper, we present a multi-wavelength (0.9-4.4 micron) morphological analysis of 33 dusty star-forming galaxies in GOODS-S at 1.5$<$\emph{z}$<$5.5 selected from the 1.1mm GOODS-ALMA 2.0 survey \citep{gomezguijarro2022}. In order to contextualize these results, we compare this sample with larger populations of star-forming and quiescent galaxies in GOODS-S at similar redshifts and stellar masses. In Section \ref{sec:data}, we describe the 1.1mm-detected sample and our selection of star-forming and quiescent galaxies in the field. In Section \ref{sec:sizes}, we present our size measurements in the rest-optical and rest-NIR, as well as RGB images and measurements of Gini-M20 parameters to better understand the complex multi-wavelength morphologies of these galaxies. In Section \ref{sec:results}, we discuss the morphological evolution across this wavelength range, the physical processes that trigger star formation in DSFGs, and how the distribution of stellar emission suggests that the 1.1mm-selected sample has undergone bulge formation. Section \ref{sec:conclusion} summarizes our main findings. Throughout this work we adopt a flat $\Lambda$CDM cosmology with $\Omega_m=0.3$, $\Omega_\Lambda=0.7$, and $H_0=70$km s$^{-1}$Mpc$^{-1}$.

\section{Data}
\label{sec:data}

\subsection{GOODS-ALMA}

We select galaxies from GOODS-ALMA 2.0 \citep{gomezguijarro2022}, a 1.1mm blind survey in GOODS-S combining high \citep[$\sim$0.2-0.3\arcsec, see ][]{franco2018} and low resolution ($>$1\arcsec) ALMA data. Out of 44 sources in the catalog with S/N $>$ 5$\sigma$, we find 33 galaxies that fall in the footprint of the GOODS-S JWST/NIRCam mosaic, matching within 0.2\arcsec of sources identified in the FRESCO photometric catalog. 
We use 1.1mm dust continuum sizes from \citet{gomezguijarro2022}, which were measured by fitting a 2D Gaussian in the $uv$ plane. In addition, for sources without a spectroscopic redshift from the FRESCO survey (described below), we use spectroscopic redshifts compiled in GOODS-ALMA 2.0.  

\begin{figure}
    \centering
    \includegraphics[width=8.5cm, keepaspectratio]{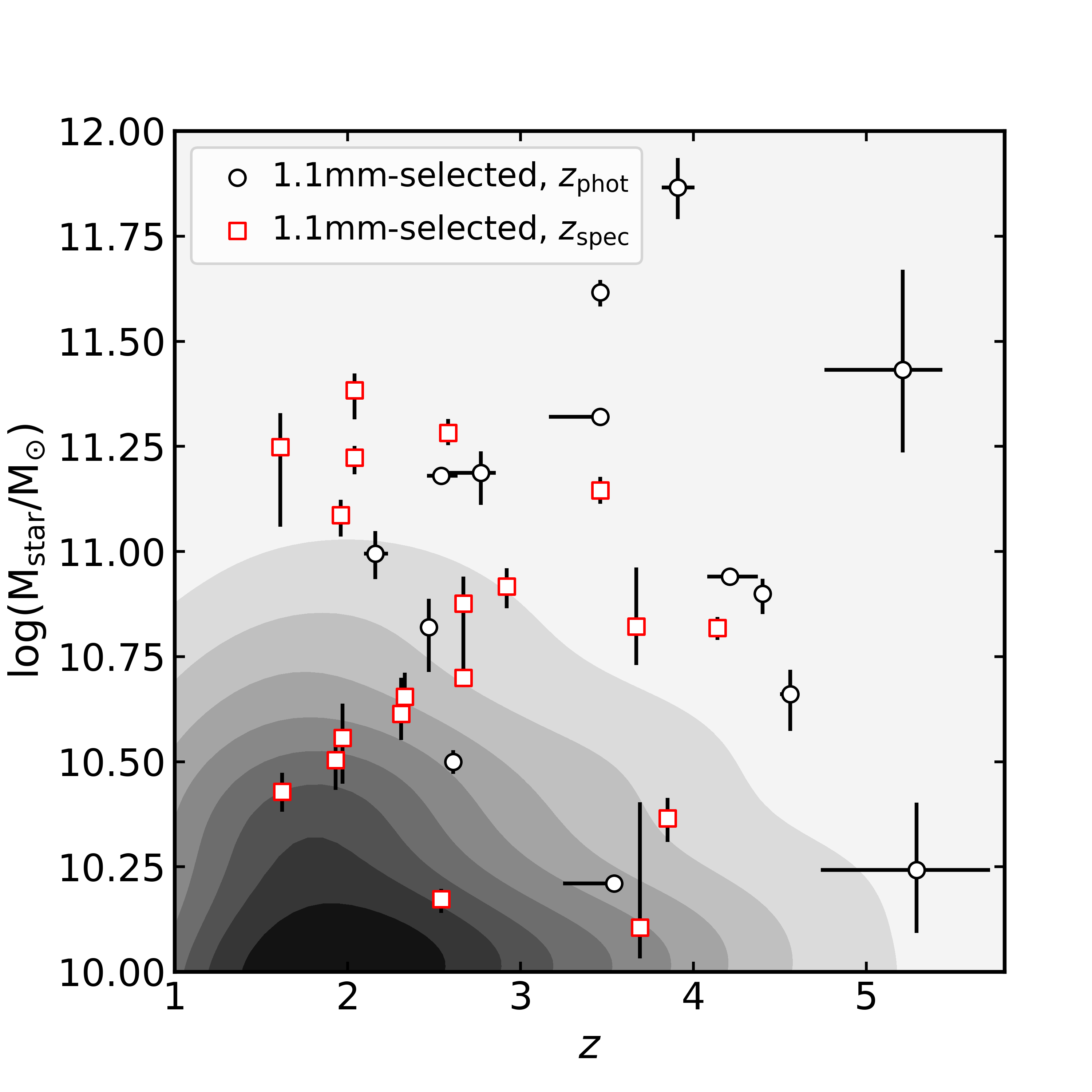}
    \caption{The stellar mass and redshift of the 1.1mm-selected GOODS-ALMA sample (white points), with field galaxies (with a stellar mass of log(M/M$_\odot) >$ 10 shown in gray contours. The mass and redshift selections for GOODS-S field galaxies is described in Section \ref{sec:nircam}.}
    \label{fig:mass_z}
\end{figure}

\subsection{NIRCam imaging}

\label{sec:nircam}

To study the rest-optical and rest-NIR morphologies of the GOODS-ALMA galaxies, we use deep imaging of GOODS-S from FRESCO \citep{oesch2023}, JADES \citep{eisenstein2023a,eisenstein2023b,rieke2023}, and JEMS \citep{williams2023} in F090W, F115W, F150W, F182M, F200W, F210M, F277W, F356W, F410M, and F444W.  These data were obtained from the Mikulski Archive for Space Telescopes (MAST) at the Space Telescope Science Institute \citep{doi_fresco, doi_jades, doi_jems}. Mosaics of these fields come from public data which have been reduced using the software package \texttt{grizli} \citep{grizli2019,grizli2021,grizli2022}. In addition, these fields are covered by HST imaging \citep{giavalisco2004,grogin2011,Koekemoer2011} with filters F435W, F606W, F775W, F814W, F850W, F105W, F125W, F140W, and F160W added to the FRESCO photometric catalogs (see below).

\begin{figure}
    \centering
    \includegraphics[width=8.5cm, keepaspectratio]{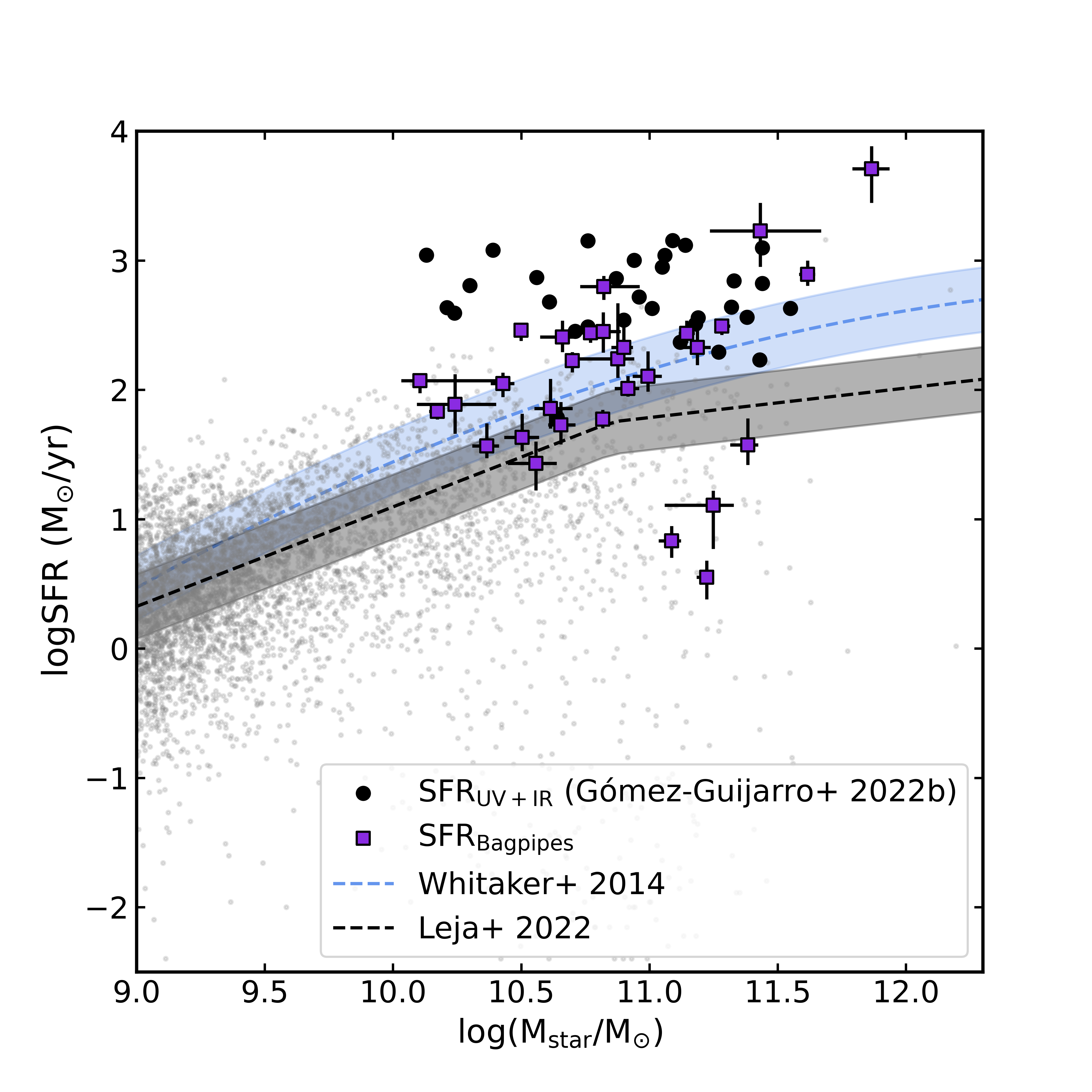}
    \caption{
    SFRs obtained through SED fitting with UV-NIR data alone (purple squares) are systematically lower at a given stellar mass than from SED fitting that incorporates the mid-IR to mm \citep[black circles,][]{gomezguijarro2022b}. We also plot the star-forming main sequence relations from \citet{leja2022} (dashed black line) and \citet{whitaker2014} (blue dashed line), where shaded regions assume 0.25 dex scatter in the main sequence, as well as GOODS-S field galaxies (gray points) at $1<z<5.5$. }
    \label{fig:sfr_mass}
\end{figure}

For the morphological analysis, we obtain priors for fluxes, axis ratios, and position angles from the GOODS-S/FRESCO photometric catalog, which was produced by running SExtractor \citep{SExtractor} and \texttt{eazy} \citep{eazy}; for further details on the construction of this catalog, see \citet{weibel2024}. Additionally, stellar masses, star formation rates, and photometric redshifts are derived using SED fitting with \texttt{bagpipes} \citep{bagpipes}. When fitting with \texttt{bagpipes}, we assume a delayed tau model SFH, a Chabrier IMF \citep{Chabrier2003} and a \citet{Calzetti2000} dust attenuation model with redshift priors from \texttt{eazy}. 

Within the 1.1mm-selected sample of 33 galaxies, the redshift ranges from \emph{z}=1.5 to \emph{z}=5.5, and stellar mass ranges from logM$_{\text{star}}$/M$_{\odot}$=10 to log M$_{\text{star}}$/M$_{\odot}$=12 (Figure \ref{fig:mass_z}). A subset of 11 galaxies have spectroscopic redshifts obtained from the FRESCO F444W/grism observations. These redshifts are all in close agreement ($\Delta z < 0.1$) with previous spectroscopic redshift measurements with the exception of the A2GS32, which has a $z_{spec}=2.04$ from a detection of Pa$\alpha$ with F444W/grism in comparison to $z_{spec}=2.251$ reported by \citet{momcheva2016} using [OII]. A further 8 galaxies have spectroscopic redshifts from previous studies (see Table \ref{tab:sizes}); in total, 19 out of 33 galaxies have $z_{spec}$. Because the \texttt{bagpipes} catalog only includes galaxies with coverage in FRESCO footprint, 29 out of the 33 galaxies in the GOODS-S mosaic are in the catalog. For the remaining four galaxies, we use stellar masses and star formation rates from \citet{gomezguijarro2022b} and photometric redshifts from SED fitting with \texttt{eazy} \citep{eazy}. We repeat all key analyses and determine that our main conclusions do not change if we remove these four galaxies.

To compare the morphologies of the mm-selected sample from GOODS-ALMA 2.0 with other massive galaxies, we select two populations with similar stellar masses and redshifts from the GOODS-S field using the \texttt{bagpipes} catalog:

\begin{enumerate}
    \item 254 star forming galaxies with logM$_{\text{star}}$/M$_{\odot}$$>$10, photometric redshifts in the range of 1$<$$z$$<$5.5, and specific star formation rates (sSFR) $>$ 0.2/t$_{U}$, where t$_{U}$ is the age of the Universe \citep{Pacifici2016}, which accounts for the evolution of the normalization in the star-forming main sequence. In addition, we adopt a rest-frame color-color selection of (\emph{U}-\emph{V})$<$1.13, (\emph{U}-\emph{V})$<$(\emph{V}-\emph{J})*0.74+0.71, and/or $(V-J)>$1.6 following \citet{leja2019}.
    \item 85 quiescent galaxies with logM$_{\text{star}}$/M$_{\odot}$$>$10, photometric redshifts in the range of 1$<$$z$$<$3, specific star formation rate (sSFR) $<$ 0.2/t$_{U}$, and rest-frame color-color selection of (\emph{U}-\emph{V})$>$1.13, (\emph{U}-\emph{V})$>$(\emph{V}-\emph{J})*0.74+0.71, and $(V-J)<$1.6.
\end{enumerate}

\begin{figure*}
    \centering
    \includegraphics[width=18cm, keepaspectratio]{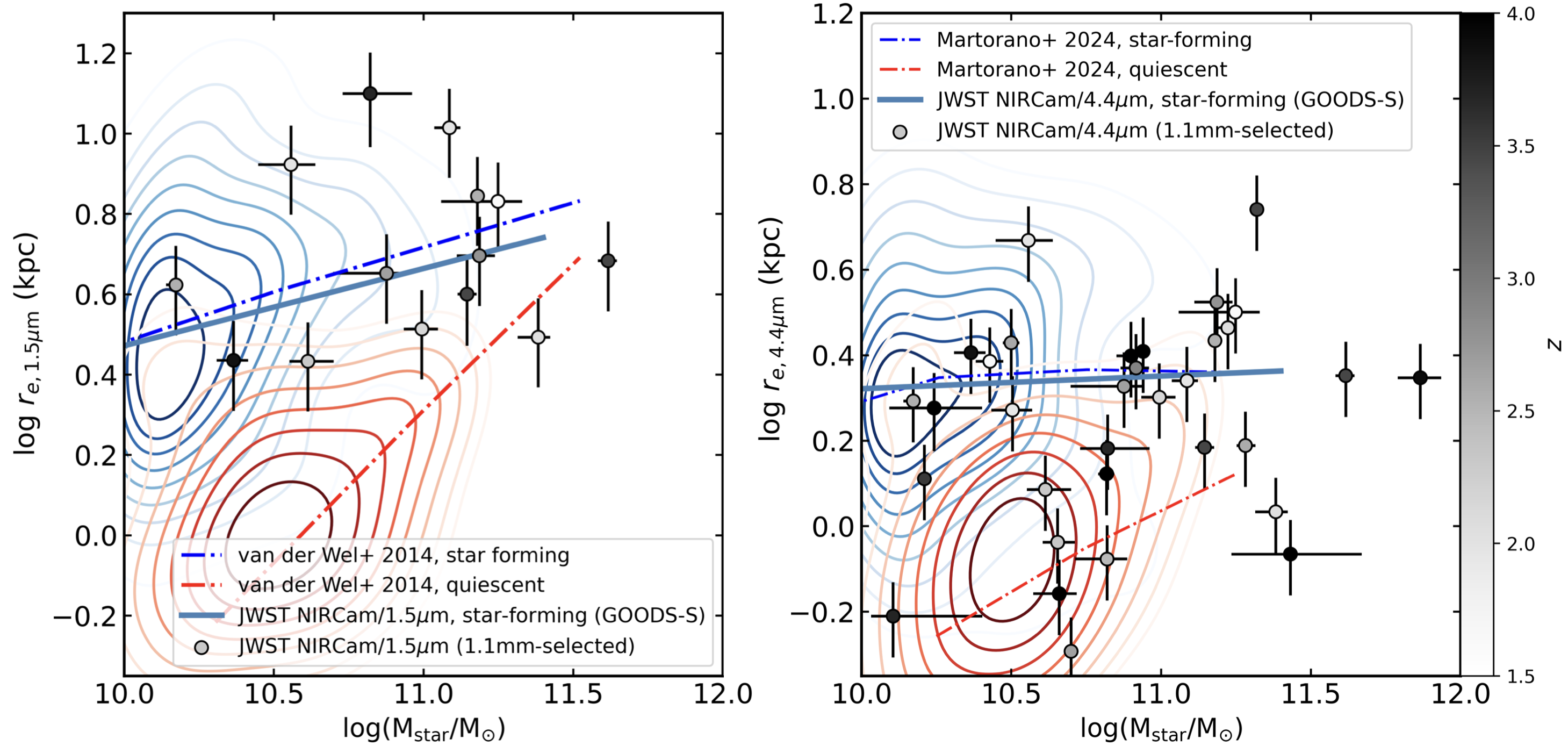}
    \caption{The effective radius along the semi-major axis at a given mass is larger in 1.5$\mu$m (left) and smaller in 4.4$\mu$m sizes (right) for 1.1mm-selected sample (gray points) relative to the field sample of star-forming galaxies described in Section \ref{sec:nircam} (blue contours). The average size-mass relation for the field sample are the solid blue lines. In 4.4$\mu$m, the 1.1mm-selected sample aligns with the quiescent population described in Section \ref{sec:nircam} (red contours). Size-mass relations from the literature in the rest-optical \citep{vanderwel2014} and rest-NIR \citep{martorano2024} are shown as blue (star-forming) and red (quiescent) dashed lines. The shading of 1.1-mm selected galaxies is based on redshift. Out of 22 galaxies that are detected in 1.5$\mu$m, the 14 points shown represent those with good fits as determined by \texttt{GALFIT}.} 
    \label{fig:size_mass}
\end{figure*}

Since we primarily use the F150W band to analyze rest-optical morphologies,  we exclude 6 quiescent galaxies at $z>3$ because F150W is redshifted into the rest-NUV. Whereas this is not a significant issue for morphological studies of star-forming galaxies \citep{ono2024}, quiescent galaxies that are no longer forming new stars are dramatically fainter in the rest-NUV relative to rest-optical wavelengths. 

Using SFRs from \citet{gomezguijarro2022b}, which are measured using FIR+UV luminosities derived from SED fitting, the majority of the 33 galaxies in our mm-selected sample are starbursts. However, SFRs measured through SED fitting with \texttt{bagpipes} \citep{bagpipes} using HST+JWST/NIRCam data are systematically lower (Figure \ref{fig:sfr_mass}), shifting $\sim$75\% of the galaxies within or below 0.25 dex of the main sequence relation from \citet{whitaker2014}. We note that while the SED based SFR-$M_*$ relation derived by \citet{leja2022} is systematically lower relative to the UV+IR SFRs presented in \citet{whitaker2014} by 0.3 dex, the discrepancy herein is a factor of two larger (0.6 dex)

The systematic offset between SFRs from \citet{gomezguijarro2022b}, where SED fitting includes mid-IR and mm data, and SFRs which are derived only from rest-optical and rest-NIR data is consistent with earlier studies \citep{elbaz2018, xiao2023} of the GOODS-ALMA sample which find that optically thick dust in these galaxies breaks the assumption of energy balance between the UV and IR. This discrepancy illustrates that SED fitting of rest-optical and rest-NIR data alone can lead to underestimates in SFR compared to methods that directly combine the observable unobscured and obscured star formation. \citet{buat2019} similarly find that SED fitting with the stellar continuum alone in dust-rich galaxies can lead to underestimating SFRs by as much as a factor of 10.

\section{Morphological Analysis}

\label{sec:sizes}

\subsection{Size measurements}

\label{sec:sersic}

\begin{table*}[t]
    \centering
    \fontsize{8}{9}
    \selectfont
    \begin{tabular}
    {|ccccccccc|}
        \hline
       ID  & $\alpha$ (J2000) & $\delta$ (J2000) & log(M$_\mathrm{*}$/M$_\odot$) & \emph{z} & r$_{e, 1.5\mu \mathrm{m}}$ & r$_{e, 4.4\mu \mathrm{m}}$ & $n_{s, 1.5\mu \mathrm{m}}$ & $n_{s, 4.4\mu \mathrm{m}}$ \\
        & (deg) & (deg) & & & (kpc) & (kpc) &  & 
       \\ \hhline{|=========|}
        A2GS1 & 53.118813 & -27.782884 & 10.61$_{-0.06}^{+0.09}$ & 2.31$^{*, \text{†}}$ & 2.7 $\pm$ 0.7 & 1.2 $\pm$ 0.2 & 0.29 $\pm$ 0.09 & 1.3 $\pm$ 0.3\\
        A2GS2 & 53.142840 & -27.827895 & 11.14$_{-0.03}^{+0.03}$ & 3.46$^{\text{†}}$ & 4 $\pm$ 1 & 1.5 $\pm$ 0.3 & 4 $\pm$ 1 & 3.1 $\pm$ 0.6\\
        A2GS3 & 53.148870 & -27.821185 & 11.28$_{-0.03}^{+0.03}$ & 2.58$^{*, \text{†}}$ & -- & 1.5 $\pm$ 0.3 & -- & 2.2 $\pm$ 0.4\\
        A2GS4 & 53.063887 & -27.843804 & 10.91$_{-0.05}^{+0.04}$ & 2.92$^{\text{†}}$ & -- & 2.3 $\pm$ 0.5 & -- & 1.3 $\pm$ 0.3\\
        A2GS6 &  53.158397 & -27.733588 & 10.8$_{-0.09}^{+0.01}$ & 3.67$^{*}$ & 13 $\pm$ 3 & 1.5 $\pm$ 0.3 & 2.9 $\pm$ 0.9 & 1.9 $\pm$ 0.4\\
        A2GS8 &  53.082752 & -27.866560 & 11.61$_{-0.03}^{+0.03}$ & 3.46$_{-0.01}^{+0.02}$ & 5 $\pm$ 1 & 2.3 $\pm$ 0.5  & 1.0 $\pm$ 0.3 & 2.5 $\pm$ 0.5\\
        A2GS9 & 53.181383 & -27.777572 & 10.9$_{-0.2}^{+0.06}$ & 2.67$^{\text{†}}$ & 4 $\pm$ 1 & 2.1 $\pm$ 0.4 & 1.1 $\pm$ 0.3 & 0.8 $\pm$ 0.2\\
        A2GS11 & 53.082069 & -27.767279 & 11.86$_{-0.07}^{+0.07}$ & 3.91$_{-0.09}^{+0.09}$ & -- & 2.2 $\pm$ 0.4 & -- & 0.7 $\pm$ 0.1\\
        A2GS12 & 53.092818 & -27.801328 & 10.36$_{-0.06}^{+0.05}$ & 3.85$^{\text{†}}$ & 2.7 $\pm$ 0.7 & 2.6 $\pm$ 0.5  & 0.8 $\pm$ 0.2 & 0.8 $\pm$ 0.2\\
        A2GS13 &  53.157199 & -27.833492 & 10.42$_{-0.05}^{+0.05}$ & 1.62$^{*,\text{†}}$ & -- & 2.4 $\pm$ 0.5 & -- & 1.3 $\pm$ 0.3\\
        A2GS14 &  53.071752 & -27.843698 & 11.09$_{-0.05}^{+0.04}$ & 1.96$^{\text{†}}$ & 10 $\pm$ 3 & 2.2 $\pm$ 0.4 & 4 $\pm$ 1 & 2.7 $\pm$ 0.5\\
        A2GS15 &  53.108810 & -27.869037 & 10.90$_{-0.05}^{+0.04}$ & 4.40$_{-0.01}^{+0.01}$ & -- & 2.6 $\pm$ 0.5  & -- & 0.44 $\pm$ 0.09\\
        A2GS17 &  53.183697 & -27.836500 & 10.2$_{-0.15}^{+0.16}$ & 5.3$_{-0.6}^{+0.4}$ & -- & 1.9 $\pm$ 0.4 & -- & 1.2 $\pm$ 0.3 \\
        A2GS18 & 53.070260 & -27.845595 & 10.1$_{-0.07}^{+0.3}$ & 3.69$^{*,\text{†}}$ & -- & 0.6 $\pm$ 0.1 & -- & 3.5 $\pm$ 0.7\\
        A2GS21 &  53.183469 & -27.776661 & 10.70$_{-0.02}^{+0.02}$ & 2.67$^{\text{†}}$ & -- & 0.6 $\pm$ 0.1 & -- & 2.7 $\pm$ 0.5\\
        A2GS22 &  53.092372 & -27.826850 & 10.50$_{-0.03}^{+0.03}$ & 2.61$_{-0.03}^{+0.03}$ & -- & 2.7 $\pm$ 0.5 & -- & 1.7 $\pm$ 0.3\\
        A2GS23 & 53.121858 & -27.752778 & 10.8$_{-0.1}^{+0.07}$ & 2.47$_{-0.02}^{+0.04}$ & -- & 0.8 $\pm$ 0.2 & -- & 1.4 $\pm$ 0.3\\
        A2GS24 &  53.092391 & -27.803269 & 11.00$_{-0.06}^{+0.05}$ & 3.23$_{-0.07}^{+0.07}$ & 3 $\pm$ 0.8 & 2.0 $\pm$ 0.4 & 0.4 $\pm$ 0.1 & 1.5 $\pm$ 0.3\\
        A2GS25 &  53.160620 & -27.776287 & 10.17$_{-0.03}^{+0.02}$ & 2.54$^{*,\text{†}}$ & 4 $\pm$ 1 & 2.0 $\pm$ 0.4 & 4 $\pm$ 1 & 0.7 $\pm$ 0.1\\
        A2GS26 & 53.090782 & -27.782492 & 10.50 $_{-0.07}^{+0.07}$ & 1.93$^{\text{†}}$ & -- & 1.9 $\pm$ 0.4 & -- & 0.7 $\pm$ 0.1\\
        A2GS27 & 53.111595 & -27.767864 & 10.66$_{-0.09}^{+0.06}$ & 4.56$_{-0.06}^{+0.03}$ & -- & 0.7 $\pm$ 0.1 & -- & 1.7 $\pm$ 0.4\\
        A2GS28 &  53.137093 & -27.761411 & 10.6$_{-0.1}^{+0.09}$ & 1.97$^{\text{†}}$ & 8 $\pm$ 2 & 4.7 $\pm$ 0.9 & 1.1 $\pm$ 0.3 & 0.6 $\pm$ 0.1\\
        A2GS29 & 53.087184 & -27.840242 & 11.32 & 3.5$_{-0.3}^{+0.02}$ & -- & 6 $\pm$ 1 & -- & 2.2 $\pm$ 0.4\\
        A2GS31 & 53.224499 & -27.817250 & 11.18 & 2.54$_{-0.08}^{+0.09}$ & 7 $\pm$ 2 & 2.7 $\pm$ 0.5 & 0.9 $\pm$ 0.3 & 1.0 $\pm$ 0.2\\
        A2GS32 &  53.077331 & -27.859632 & 11.22$_{-0.04}^{+0.03}$ & 2.04$^{*,a}$ & -- & 2.9 $\pm$ 0.6 & -- & 3.5 $\pm$ 0.7\\
        A2GS33 & 53.120402 & -27.742111 & 11.4$_{-0.2}^{+0.2}$ & 5.2$_{-0.5}^{+0.2}$ & -- & 0.9 $\pm$ 0.2 & -- & 1.5 $\pm$ 0.3\\
        A2GS34 &  53.131474 & -27.841396 & 11.2$_{-0.2}^{+0.09}$ & 1.61$^{*,\text{†}}$ &  7 $\pm$ 2 & 3.2 $\pm$ 0.6 & 2.6 $\pm$ 0.8 & 2.3 $\pm$ 0.5\\
        A2GS35 & 53.069006 & -27.807141 & 10.94 & 4.2$_{-0.1}^{+0.2}$ & -- & 2.6 $\pm$ 0.5 & -- & 0.7 $\pm$ 0.1\\
        A2GS36 &  53.086635 & -27.810257  & 11.19$_{-0.08}^{+0.05}$ & 2.8$_{-0.2}^{+0.1}$ & 5 $\pm$ 1 & 3.3 $\pm$ 0.7 & 0.20 $\pm$ 0.06 & 0.9 $\pm$ 0.2\\
        A2GS39 &  53.091617 & -27.853421 & 11.38$_{-0.07}^{+0.04}$ & 2.04$^*$ & 3.1 $\pm$ 0.8 & 1.1 $\pm$ 0.2 & 0.6 $\pm$ 0.2 & 1.3 $\pm$ 0.3\\
        A2GS40 & 53.196569 & -27.757065 & 10.21 & 3.5$_{-0.3}^{+0.04}$ & -- & 1.3 $\pm$ 0.3 & -- & 1.3 $\pm$ 0.4\\
        A2GS42 &  53.154440 & -27.738686 & 10.65$_{-0.05}^{+0.06}$ & 2.33$^*$ & -- & 0.9 $\pm$ 0.2 & -- & 1.4 $\pm$ 0.3\\
        A2GS44 & 53.102654 & -27.860660 & 10.82$_{-0.03}^{+0.03}$ & 4.14$^*$ & -- & 1.3 $\pm$ 0.3 & -- & 2.9 $\pm$ 0.6\\
        \hline
    \end{tabular}
    \caption{IDS, right ascension, and declination from the GOODS-ALMA 2.0 survey \citep{gomezguijarro2022}. Effective radii and S\'ersic index come from the single component S\'ersic profile fits, and stellar masses and photometric redshifts come from the FRESCO photometric catalogs. For A2GS29, A2GS30, A2GS35, and A2GS40, stellar masses are from \citep{gomezguijarro2022}. $^{*}$indicates spectroscopic redshift from FRESCO grism. $^{\text{†}}$indicates spectroscopic redshifts from GOODS-ALMA 2.0 compilation. References for spectroscopic redshifts: \citet{vanzella2008, Wuyts2009,kurk2013,momcheva2016,inami2017, Decarli_2019,Zhou2020, Garilli2021}.
 $^{a}$ $z_{\text{spec}}=2.251$ from \citet{momcheva2016}.}.
    \label{tab:sizes}
\end{table*}

We perform single component S\'ersic profile fits using \texttt{GALFIT} \citep{galfit} on 6\arcsec{} by 6\arcsec{} cutouts. We obtain initial values for effective radius along the semi-major axis ($r_e$), S\'ersic index ($n_S$), magnitude, position angle, and axis ratio from the FRESCO photometric catalog. 
Parameters are constrained to within $r_e$=[0.3 pix, 400 pix], $n_S$=[0.2,6], axis ratio=[0.0001,1], and we allow the total magnitude to vary within $\pm$3 mag of its initial value. Neighboring objects are flagged and masked out using the detection segmentation map.

The sizes in F150W and F444W as well as the stellar masses and redshifts of the 1.1mm-selected galaxies are listed in Table \ref{tab:sizes}. Due to differences in NIRCam coverage, only 24 of the 33 GOODS-ALMA selected galaxies are imaged in both F150W and F444W. Out of these 24 galaxies, 14 are determined to have a reliable S\'ersic fit by \texttt{GALFIT}. Errors for sizes are determined by using Monte Carlo bootstrap methods as follows. For each galaxy, we create 100 images by adding random perturbations drawn from the error image to the science image. These 100 images are fit with \texttt{GALFIT}, and the standard deviation in the distribution of sizes is taken as the uncertainty in size for each galaxy. Because the 1.1mm-selected sample in general is quite bright (particularly in redder bands) the bootstrapped errors are relatively small (less than 1\% on average in F444W), and likely underestimate the true uncertainty in the single component S\'ersic fits, reflecting instead the brightness of the 1.1mm-selected sample relative to the background. To get errors that more realistically capture the uncertainty in our fits, we independently measure sizes using the Python package \texttt{statmorph} \citep{statmorph}. We find that the \texttt{GALFIT} and \texttt{statmorph} sizes are generally consistent, with median size ratios of 0.98 (F150W) and 0.92 (F444W), and standard deviation in the size ratios of 0.25 (F150W) and 0.2 (F444W). The standard deviations correspond to fractional systematic errors of 25\% in F150W and 20\% in F444W, which we add in quadrature to the bootstrapped errors.
 
We note that although we use F150W to trace the rest-optical and F444W to trace the rest-NIR, the wide redshift range of our 1.1mm-selected sample means that these classifications do not apply uniformly, particularly at higher redshifts. For galaxies at $z<4.5$ (30 out of 33 galaxies), F444W probes the rest-frame NIR. However, for galaxies at $z>2.75$ (16 out of 33 galaxies), F150W more accurately traces rest-frame NUV emission. This wavelength shift affects a smaller subset of the sample for which we have size measurements: only 5 out of 14 galaxies with robust \texttt{GALFIT} fits  lie at $z>2.75$.  We return to this issue in Section~\ref{sec:results} to affirm that all conclusions remain robust when repeating analyses limited to $z<2.75$.

\begin{figure}
    \centering
    \includegraphics[width=8.5cm, keepaspectratio]{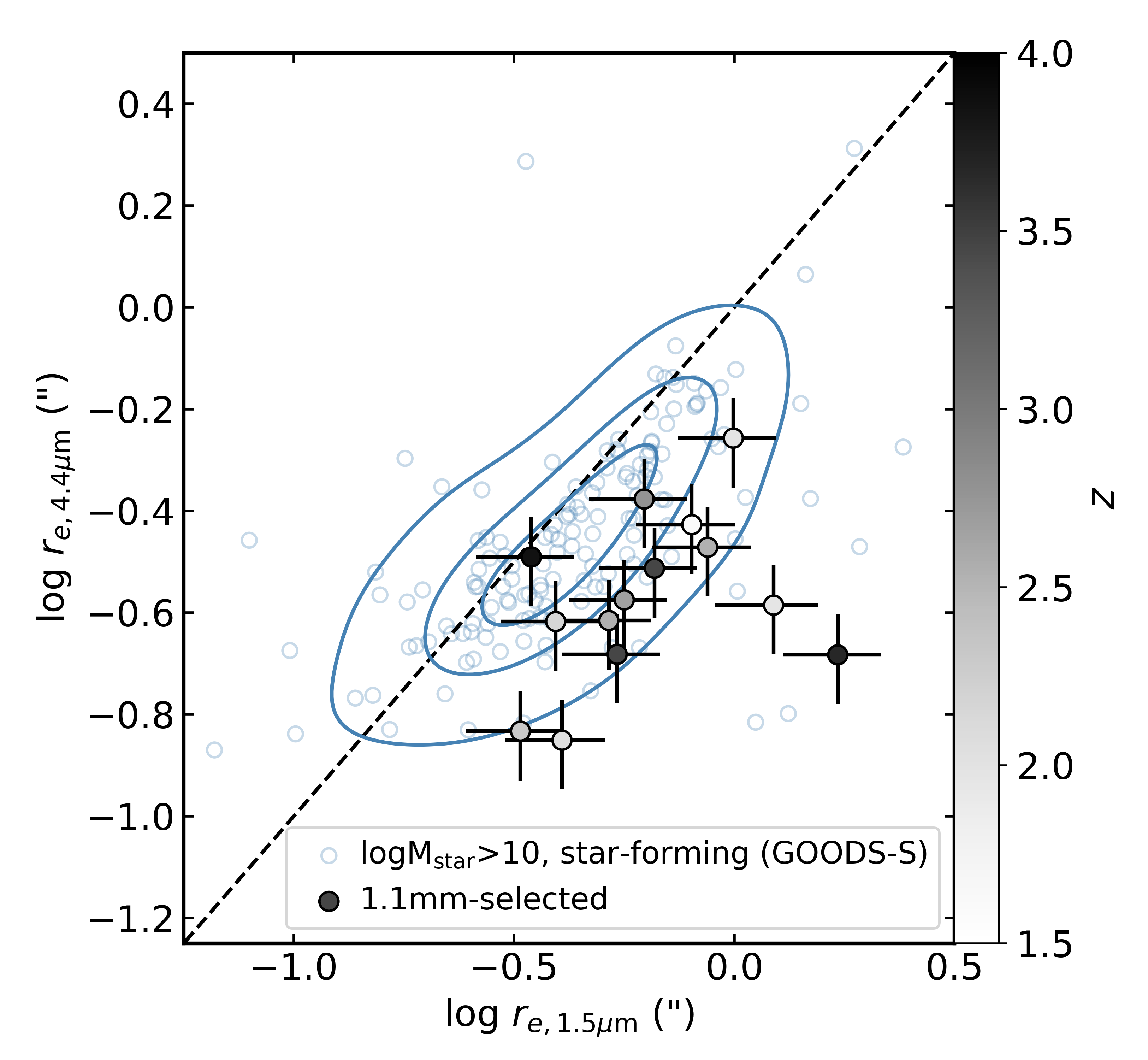}
    \caption{A direct comparison of effective radius measured at 1.5$\mu$m vs at 4.4$\mu$m shows that the 1.1-mm selected sample are unusually large in 1.5$\mu$m relative to 4.4$\mu$m. Blue contours are the 20th, 50th, and 80th percentiles of the distribution of the field sample (blue points). The shading of 1.1-mm selected galaxies is based on redshift.} 
    \label{fig:150_vs_444}
\end{figure}

In Figure \ref{fig:size_mass}, the average size-mass relation measured at wavelengths of 1.5$\mu$m and 4.4$\mu$m is shown for the field sample of star-forming galaxies in GOODS-S discussed in Section \ref{sec:nircam}. We derive our size-mass relation for star-forming galaxies by performing a linear regression on field 
star-forming galaxies with reliable structural fits, as determined by \texttt{GALFIT}. At 1.5$\mu$m, 1.1mm-selected galaxies are broadly similar to other star-forming galaxies, though three galaxies reside $>$4 kpc above the 1.5$\mu$m size-mass relation. In contrast, at 4.4$\mu$m, $\sim$60\% fall below the size-mass relation. In other words, the 1.1mm-selected galaxies are on average typical in size given their stellar mass in the rest-optical, but are relatively smaller in the rest-NIR. The 1.1mm-selected sample instead appears more closely aligned with the quiescent size-mass relation in the rest-NIR. Figure \ref{fig:150_vs_444} directly compares 1.5$\mu$m and 4.4$\mu$m sizes, showing that $\sim$70\% of the 1.1mm-selected sample falls outside the 50th percentile contour of the field star-forming galaxies. The median difference in the log-scale sizes is higher for the 1.1mm-selected sample, with a median value of $\log R_{e, 1.5_{\mu \mathrm{m}}}$ 0.17 dex larger than $\log R_{e, 4.4_{\mu \mathrm{m}}}$ for the 1.1mm-selected sample, compared to an equivalent offset of only 0.03 dex for field star-forming galaxies.

\begin{figure*}[h]
    \centering
    \includegraphics[width=16cm, keepaspectratio]{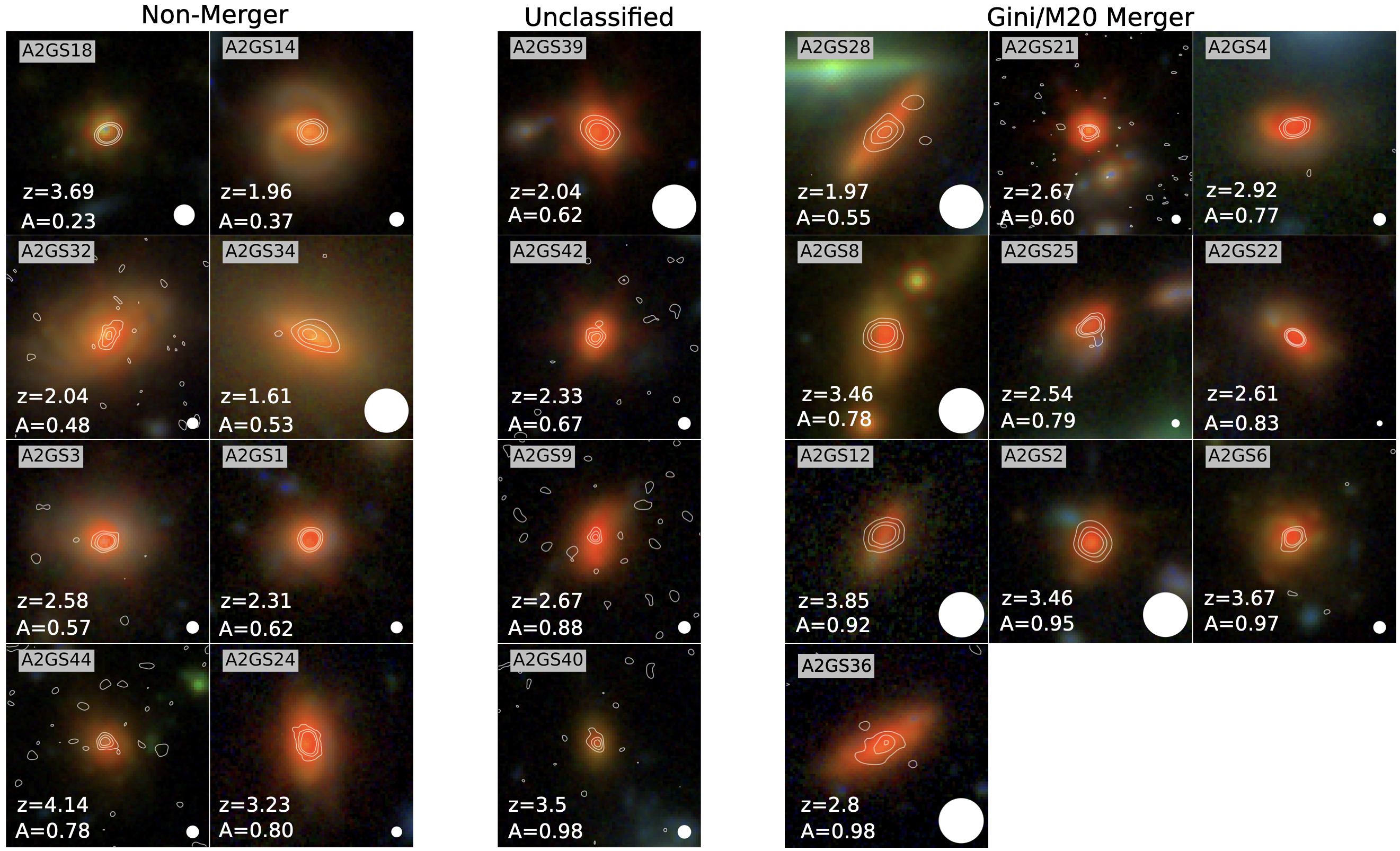}
    \caption{Bright red cores are visible in 3.2$\arcsec$ by 3.2$\arcsec$  RGB stamps of the ALMA-selected sample, where the false color images combine the 1.5$\mu$m, 2.7$\mu$m, and 4.4$\mu$m NIRCam bands with 2$\sigma$, 4$\sigma$, and 6$\sigma$ contours from ALMA mm observations in white. The ALMA beam size is shown in the bottom right corner of each stamp. Galaxies are separated into three groups depending on whether they are not classified as a merger (left), their merger status is ambiguous (center), or they are classified as a merger according to their Gini/M20 parameters in 1.5$\mu$m \citep{Lotz2004,Lotz2008}. Within these groups, galaxies are sorted from the top down by increasing asymmetry parameter.}
    \label{fig:rgb}
\end{figure*}

\begin{figure*}
    \centering
    \includegraphics[width=16.5cm, keepaspectratio]{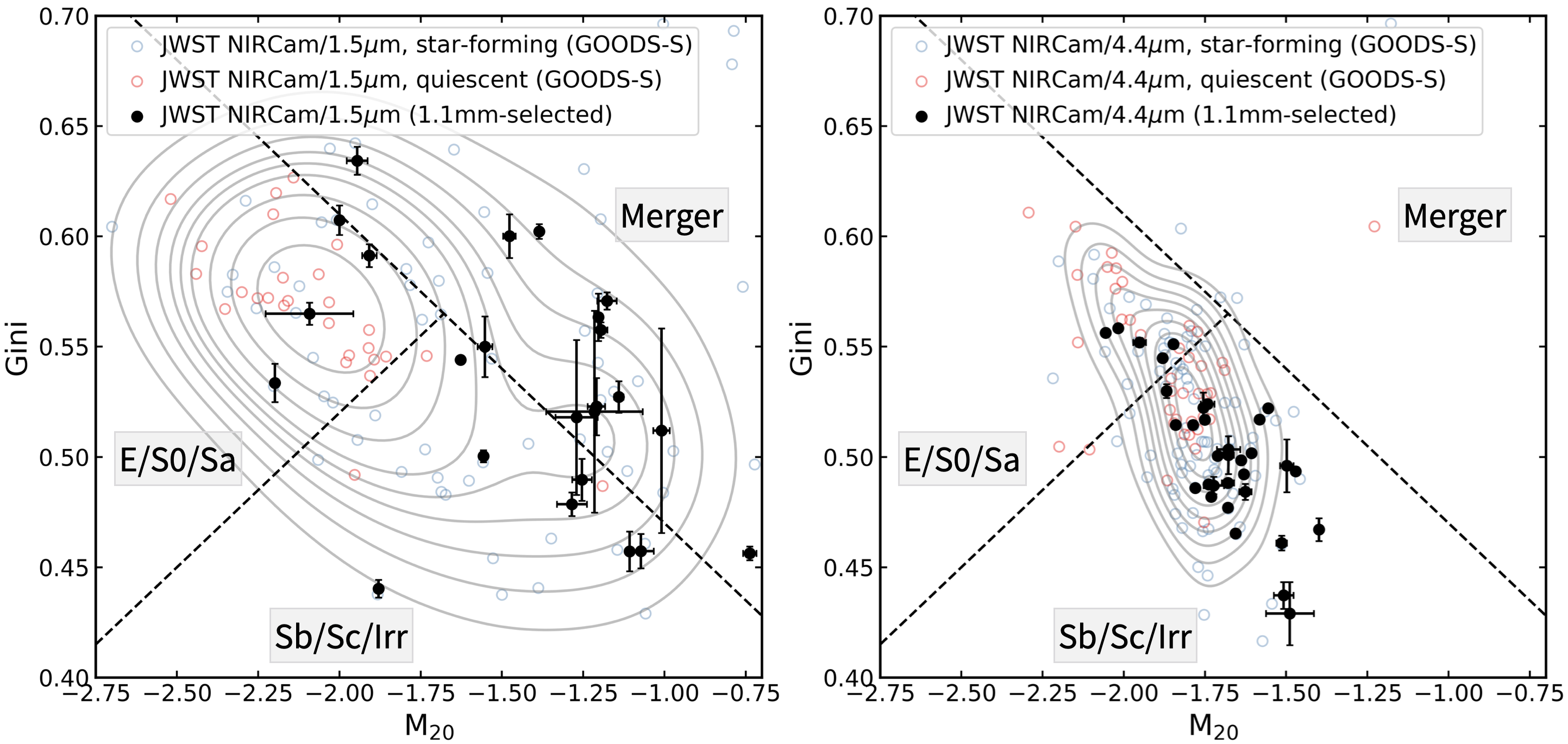}
    \caption{Left: Using the classification scheme from \citet{Lotz2008}, the Gini/M20 parameters at 1.5$\mu$m indicate that about half the star-forming field sample (blue circles) and the 1.1mm-selected sample are mergers. Most quiescent galaxies (red circles) fall into the elliptical wedge of the plot. Right: At 4.4$\mu$m, only a handful of massive galaxies are classified as mergers, and all of the 1.1mm-selected galaxies are classified as non-mergers. The distribution of Gini-M20 values as a function of wavelength is discussed in more detail in Section \ref{sec:mergers}} 
    \label{fig:giniM20}
\end{figure*}

The compact rest-NIR sizes of the 1.1mm-selected sample are neither a resolution nor a redshift effect. In order to rule out that the difference in resolution between shorter and longer wavelength NIRCam bands is driving this trend, we compare the sizes in the PSF-homogenized 1.5$\mu$m imaging to the F444W PSF. We find that the recovered sizes in the 1.5$\mu$m imaging are on average only 1\% smaller and thus conclude that the difference in resolution has a minimal effect on our results. 
 Moreover, the color-coding of the symbols in Figures \ref{fig:size_mass} and \ref{fig:150_vs_444} demonstrate that we find no obvious correlation between rest-NIR and rest-optical sizes with redshift despite the 1.1mm-selected sample having a the slightly higher median redshift relative to the field star-forming sample.
 This discrepancy in size across the rest-optical to rest-NIR is further explored in Section \ref{sec:size-wavelength}.

\subsection{Non-parametric morphologies}
\label{sec:morph}

We create RGB images of the 1.1mm-selected sample by combining 1.5$\mu$m, 2.7$\mu$m, and 4.4$\mu$m imaging using the \citet{lupton2004} scheme (Figure \ref{fig:rgb}) for galaxies that are detected in F150W. Additionally, we add ALMA contours of dust continuum emission. The dust emission appears to be co-spatial with 4.4$\mu$m emission for the majority of our sample.  The underlying structure of dust, which may be clumpy, is only marginally resolved. In several cases, there are noticeable offsets between 1.1mm dust emission and the center of the galaxy as traced by stellar emission. Dusty star formation in these galaxies appears to be decoupled from the previous epochs of central star formation, with the most logical conclusion being that we are instead tracing a clump in the disk.  That said, an off-center clump may instead be a dusty merging galaxy in some cases \citep[e.g.,][]{hodge2025}, which are most prominent at mm-wavelengths and may not even be detected in sensitive JWST imaging \citep[e.g.,][]{McKinney2025, Manning2025}. Figure \ref{fig:rgb} demonstrates the morphological diversity of the 1.1mm-selected sample, which range from spiral galaxies with central bulges, to irregular systems undergoing merger events, and compact red objects. Two of the 1.1mm-selected galaxies (A2GS15 and A2GS29) are undetected in 1.5$\mu$m imaging \citep[these galaxies are discussed in more detail in][]{Zhou2020, xiao2023}.

A significant fraction of the sample has a red compact region with more extended and clumpier emission apparent in shorter wavelength bands. These RGB images are consistent with a model in which the 1.1mm-selected galaxies contain a star-forming core that is heavily obscured by dust, leading to more extended light profiles at rest-optical wavelengths relative to the rest-NIR. \citet{gillman2024} and \citet{boogaard2024} find similarly complex morphologies in shorter wavelength bands, which they note could be the product of attenuation from dust structures as well as underlying clumpy star formation.

To quantify the complexity of the clumpy emission observed in shorter wavelength bands, we measure Gini/M20 values of 1.5$\mu$m imaging \texttt{statmorph}, classifying these systems as mergers as defined in \citet{Lotz2004,Lotz2008}. Figure \ref{fig:giniM20} divides up the sample according to Gini/M20 merger classification, and also includes the asymmetry parameter. Systems classified as mergers by Gini-M20 have an average asymmetry value of 0.78$\pm 0.13$, compared to 0.55 $\pm 0.18$ for non-mergers, where errors are the standard deviation. 

Figure \ref{fig:giniM20} shows where the 1.1mm-selected sample falls in the Gini-M20 parameter space at both 1.5$\mu$m and 4.4$\mu$m. 
At 1.5$\mu$m, we find a high merger fraction of 54\% for the 1.1mm-selected sample using the Gini/M20 classification, which is comparable to a 49\% merger fraction for the field star-forming population at the same wavelength. These merger fractions are atypically high relative to other studies \citep[see ][]{stott2013, ventou2019, gillman2024, mckay2025, ren2025}. However, at 4.4$\mu$m, both the 1.1mm-selected and field samples have much lower merger fractions (0\% and 6\%, respectively), and exhibit less scatter in Gini-M20.  The drop in merger fraction at longer wavelengths is consistent with \citet{gillman2024}, who find a somewhat more modest decrease from 40\% at 1.5$\mu$m to 8\% at 4.4$\mu$m with a 2.4$\times$ larger sample. \citet{constantin2025} also find that in a sample of massive galaxies at $3<z<5$, the rest-NIR (F560W) morphologies look smooth, compact, and disk-like, with all galaxies appearing to have disk-like morphologies using the \citet{Lotz2008} Gini-M20 classification in the rest-NIR. The wavelength dependence of merger classification is discussed in greater detail in Section \ref{sec:mergers}.

\section{Discussion}

\label{sec:results}

\subsection{A steeper size-wavelength gradient for dusty galaxies}

\label{sec:size-wavelength}

\begin{figure}
    \centering
    \includegraphics[width=8.5cm, keepaspectratio]{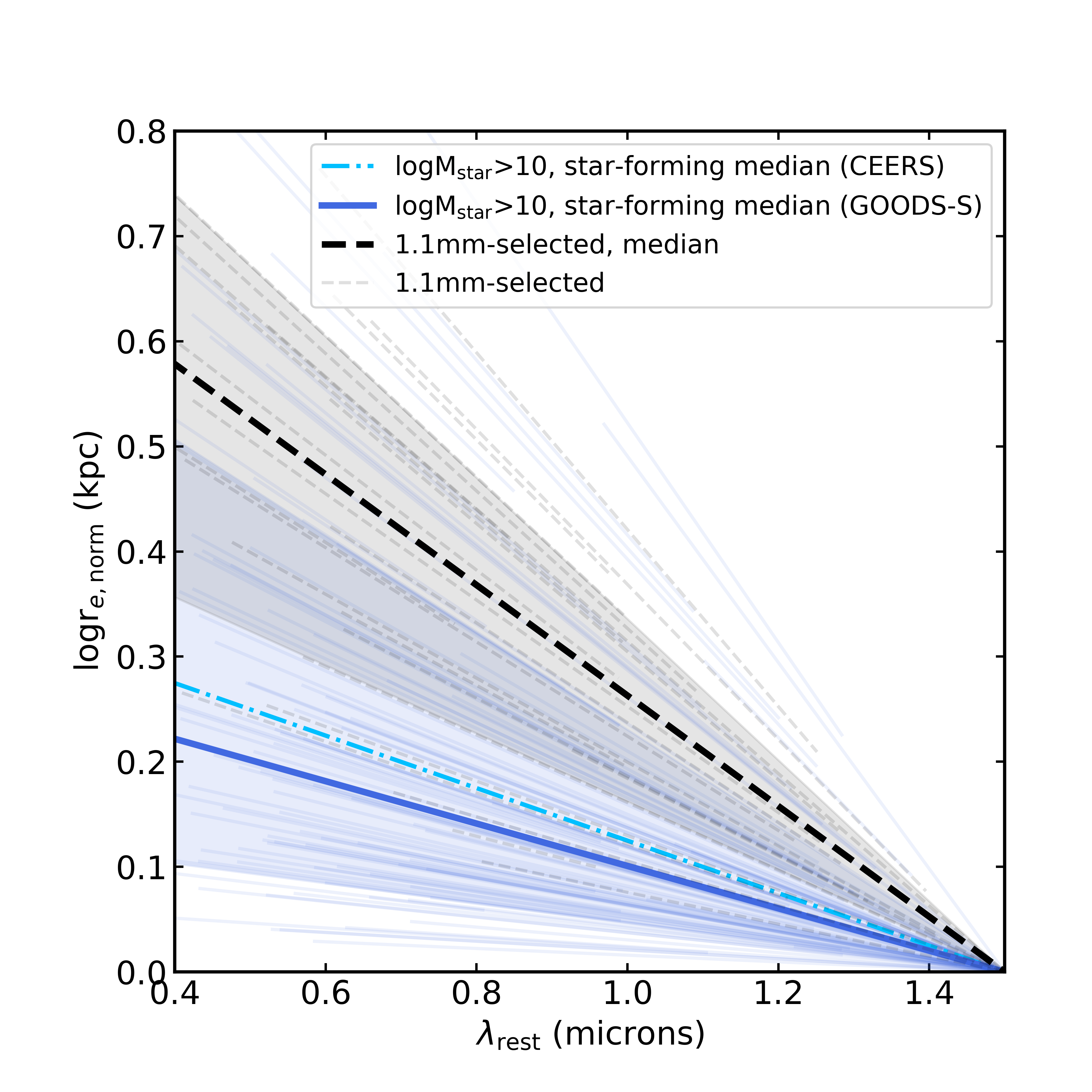}
    \caption{Lines of best fit for the normalized effective radius vs rest wavelength. The slope of this relationship is steeper for the 1.1mm-selected sample (gray dashed lines, black line is the median trend) relative to the field sample (blue lines) and a mass-selected sample from the CEERS survey \citep[cyan dashed line;][]{suess2022}. The shaded regions represent the 16th and 84th percentiles of the lines of best fit.}
    \label{fig:size_wavelength}
\end{figure}

\begin{figure*}
    \centering
    \includegraphics[width=17cm, keepaspectratio]{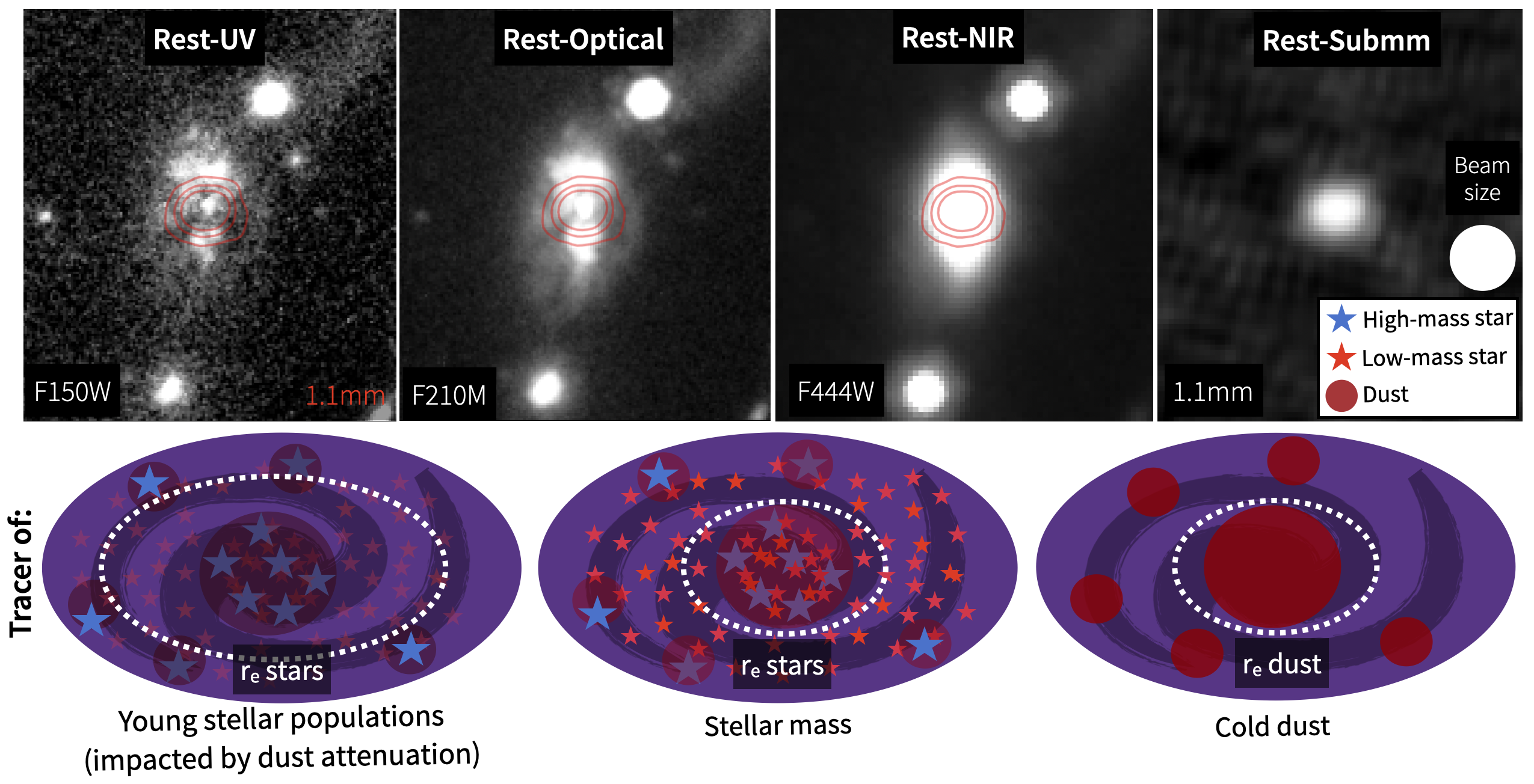}
    \caption{An illustration explaining how morphology may vary with wavelength in the galaxy A2GS8, with JWST/NIRCam imaging (top panel) and a cartoon representing the stellar populations and dust within the galaxy (bottom panel). Lower mass stars are represented by red stars, whereas massive, short-lived stars are larger blue stars. In the rest-UV (F150W) and rest-optical (F210M), we see clumpy structures that trace young stellar populations obscured by dust. In contrast, the rest-NIR (F444W) shows smooth and centrally concentrated light from stars across the IMF. In the rest-submm (1.1mm) we see compact and also centrally concentrated emission from cold dust. There may be clumpy dust structures associated with star-forming regions that are unresolved. The dashed white ellipse in the cartoon represents the effective radius in each wavelength; the effective radius of the rest-optical (and rest-UV) light from younger stars is more extended than in the rest-NIR (tracing low mass stars dominating the stellar mass budget), as well the rest-submm (tracing cold dust).  This effect is driven by the presence of concentrated dust that preferentially suppresses the central rest-UV and rest-optical light.  Rest-NIR light is less sensitive to outshining and dust effects, and thus the light distribution is more similar to that of the cold dust produced via new star formation.}
    \label{fig:cartoon}
\end{figure*}

In Section \ref{sec:sizes}, we show that while the 1.1mm-selected sample is mostly similar in size to other star-forming galaxies in the rest-optical, these same galaxies are compact at rest-NIR wavelengths. In order to explore this trend in more detail, we determine the relationship between wavelength and effective radius by leveraging all available NIRCam filters. For each galaxy in both the field population of massive star-forming galaxies and the 1.1mm-selected sample, we normalize by the size at rest-frame 1.5$\mu$m to counteract the dependence of effective radius on stellar mass and redshift, which will contribute to scatter in this relationship. We then find the line of best fit for rest wavelength versus the 1.5$\mu$m-normalized effective radius. We only consider galaxies where the Pearson correlation coefficient between wavelength and size is greater than 0.6 and where there are three or more NIRCam bands with good size measurements (i.e. they have not been flagged by \texttt{GALFIT}). 
Figure \ref{fig:size_wavelength} shows the size-wavelength gradient for the 1.1mm-selected galaxies as compared to the field population. The dusty, 1.1mm-selected galaxies have slopes that are 2.6$\times$ steeper relative to the field population. Only two galaxies in 1.1mm-selected sample have slopes that are flatter than the average size-wavelength relationship of massive star-forming galaxies in GOODS-S. As noted in Sec. \ref{sec:sizes}, the difference in resolution between 1.5$\mu$m and 4.4$\mu$m has a minimal impact on the observed size-wavelength gradient. Our result is consistent with other multi-wavelength analyses of submillimeter galaxies (SMGs) \citep[e.g.][]{chen2022,hodge2025,price2025}, which also show a strong color gradient across rest-optical to rest-NIR sizes. We also compare to \cite{suess2022}, using the 1.5$\mu$m and 4.4$\mu$m size measurements from imaging taken by CEERS program \citep{finkelstein2023} in the EGS field and confirm that this size-wavelength relationship closely matches the field population from GOODS-S.

Given that GOODS-ALMA galaxies are selected because they are detected at 1.1mm, the dust in these galaxies is likely responsible for the difference in the size-wavelength gradient. In RGB images shown in Figure  \ref{fig:rgb}, we see that many of galaxies contain a heavily dust obscured, star-forming core. Figure \ref{fig:cartoon} further illustrates an example of how morphology varies from the rest-optical to the rest-submm. Attenuation near central regions, traced by compact dust emission at rest-submm wavelengths, will dim the center at rest-optical wavelengths whereas the rest-NIR will not suffer from the effects of attenuation as strongly. This central dust attenuation results in more extended half-light radii and clumpy morphologies in the rest-optical relative to the rest-NIR. As noted in Section \ref{sec:sizes}, for the higher redshift subset of the 1.1mm-selected sample, it is more appropriate to describe observed 1.5$\mu$m emission as tracing the rest-NUV. The physical interpretation is not significantly different for these galaxies; in this case, shorter rest-wavelength emission will be more heavily biased towards younger stellar populations and impacted by dust attenuation. If we exclude galaxies at $z>2.75$ from this analysis, we find that the size-wavelength gradient is only 2\% shallower; therefore, we conclude the the wide redshift range of our 1.1mm-selected sample does not strongly impact our results. In Section \ref{sec:mergers}, we consider the physical origin of the dust and the clumpy, extended rest-optical morphologies observed in the sample of dusty star-forming galaxies.

\subsection{The spatial extent of dust relative to stellar mass}

\begin{figure}
    \centering
    \includegraphics[width=8.5cm, keepaspectratio]{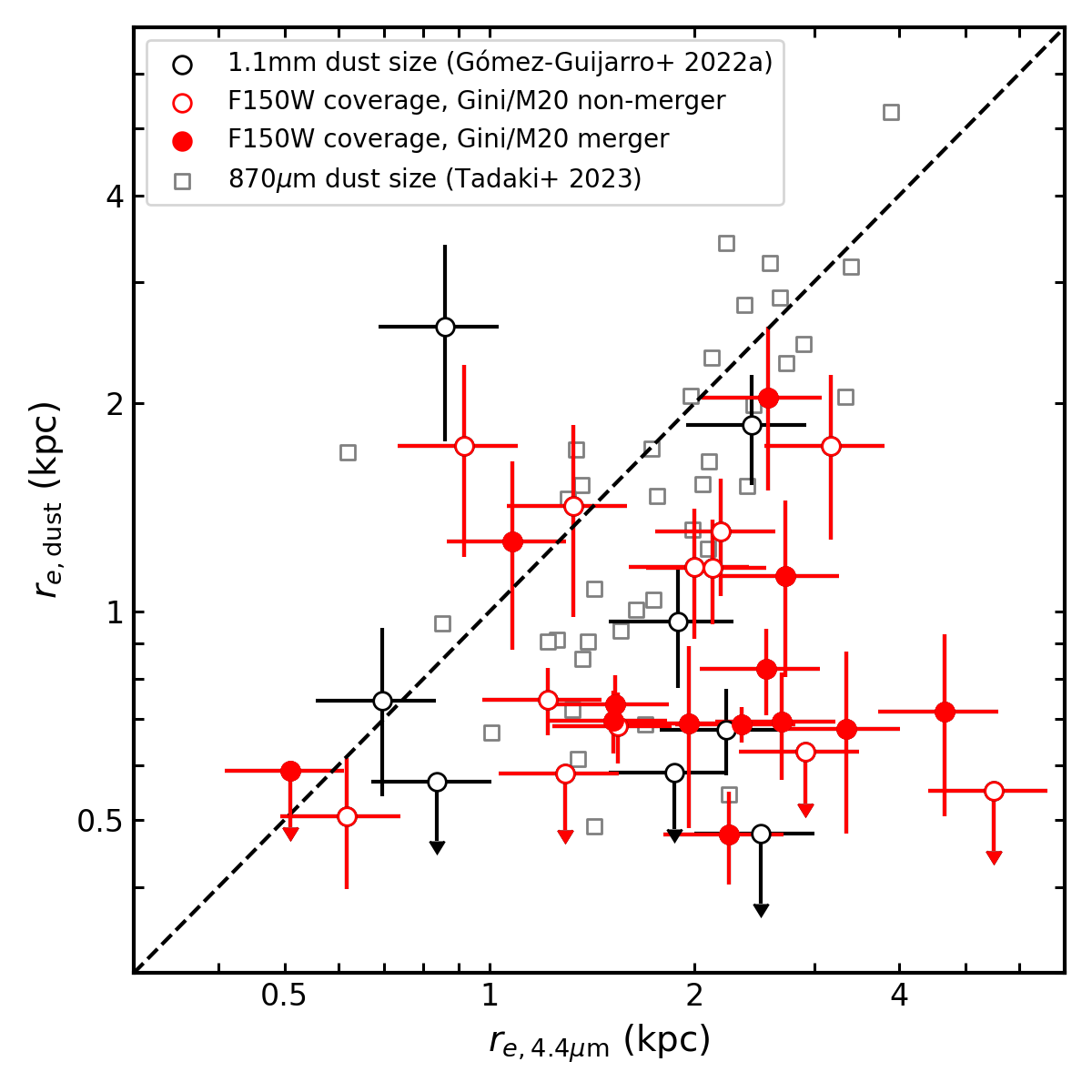}
    \caption{1.1mm dust continuum sizes tend to be more compact than 4.4$\mu$m sizes for the 1.1mm-selected sample (circular points). 870$\mu$m dust sizes from \citet{tadaki2023} are shown as gray squares. For 1.1mm-selected galaxies with 1.5$\mu$m coverage (red points), merger status from Gini/M20 is color coded, where solid points are classified as mergers and open points are non-mergers.}
    \label{fig:dust_vs_444}
\end{figure}

As noted in Section \ref{sec:morph}, a compact dusty core appears to be largely co-spatial with rest-NIR emission in the 1.1mm-selected sample, and this dust is likely the primary driver of the steep size-wavelength gradient discussed in \ref{sec:size-wavelength}. By directly comparing the extent of dust emission with the rest-NIR in Figure \ref{fig:dust_vs_444}, we test how the distribution of recent star formation (as traced by dust) compares to the bulk stellar mass \citep[as traced by the rest-NIR][]{suess2022,clausen2025}. Similar to \citet{chen2022,tadaki2023, hodge2025}, Figure \ref{fig:dust_vs_444} shows that the rest-NIR emission is typically more extended relative to the dust continuum, with 82\% of the sample having more compact 1.1mm sizes.  We note that we use the 1.1mm dust continuum size, which is sensitive to the bulk of the dust mass \citep{Scoville2016}, whereas \citet{chen2022,tadaki2023,hodge2025} all use 870$\mu$m, which traces somewhat warmer dust and is predicted to be more compact \citep{cochrane2019, popping2022} relative to 1.1mm emission. Similar to studies using 870$\mu$m, our results are in conflict with predictions from hydrodynamical simulations that the dust emission should be more extended than the stellar mass \citep{cochrane2019,popping2022}. If we only consider systems that are classified as mergers in 1.5$\mu$m via the Gini-M20 diagnostic, the median difference between 4.4$\mu$m and 1.1mm sizes is 1.6 kpc; for non-mergers, this difference is 0.9 kpc. In other words, galaxies identified as mergers in the rest-optical generally have rest-NIR sizes that are more extended relative to their 1.1mm dust emission. This suggests that these galaxies may be in an evolutionary stage in which dust traces active star formation in a compact core (potentially fueled by a merger), while the rest-NIR traces the older, bulk generation of star formation (in addition to younger dust-obscured stellar populations).

\subsection{The origin of enhanced star formation in the 1.1mm-selected sample}

\label{sec:mergers}

There are several scenarios in which dust could preferentially cause the half-light radii in the rest-optical to appear more extended relative to the rest-NIR. As discussed in the previous section, cases where rest-submm emission is more compact relative to emission at both rest-optical and rest-NIR wavelengths may be a consequence of a merger that drives central star formation. Similarly, minor mergers and accretion may instead more subtly enhance star formation in the outskirts \citep[e.g.,][]{Hill2017}. Finally, the accretion of cold gas (either continuously or episodically) may similarly result in enhanced star formation and thus dust production.  

\begin{figure}
    \centering
    \includegraphics[width=8.5cm, keepaspectratio]{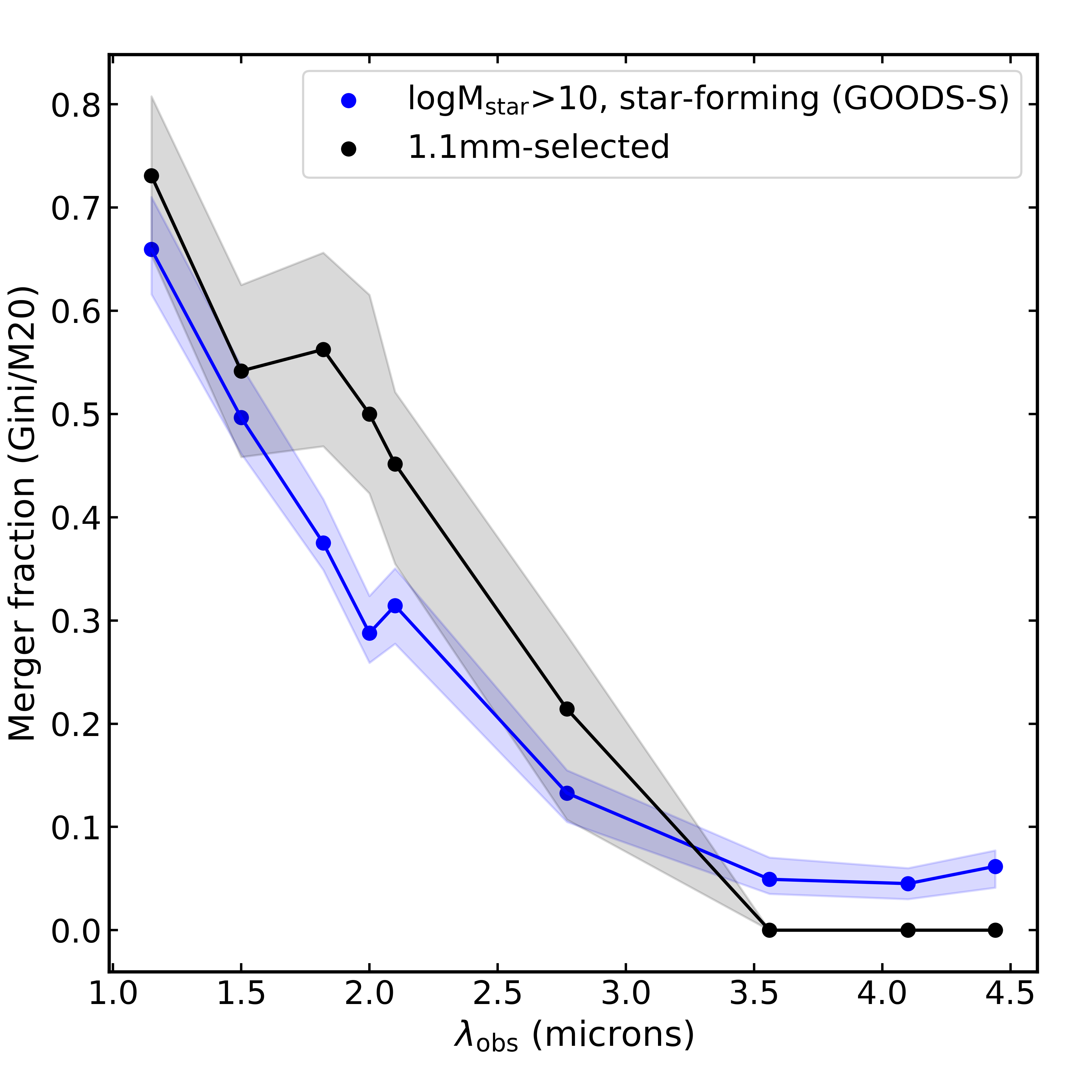}
    \caption{Classical Gini-M20 diagnostics fail to produce consistent results in merger classification across the rest-UV to rest-NIR, given that the fraction of mergers from the \citet{Lotz2008} relation drops sharply as a function of wavelength from 1.15$\mu$m to 4.4$\mu$m in both the 1.1mm-selected sample (black line) and more typical star-forming galaxies (blue line). The shaded regions represent the 16th and 84th percentiles of a bootstrapped distribution.  The two key implications include that (1) we cannot robustly conclude if 1.1mm-selected galaxies have higher merger rates than field star-forming galaxies, and (2) the merger fraction may be underestimated at longer wavelengths in particular, as visual inspection suggest that at least a few of the 1.1mm-selected galaxies are mergers.}
    \label{fig:merger_vs_wavelength}
\end{figure}

In an attempt to distinguish these various scenarios, we first consider whether a higher prevalence of major mergers within our 1.1mm-selected sample could be a potential cause of enhanced star formation. As discussed in Section \ref{sec:morph}, using Gini-M20 parameters we find similar merger fractions in the 1.1mm-selected sample and field star-forming galaxies, but significant variation in merger fraction at different wavelengths. Figure \ref{fig:merger_vs_wavelength} illustrates the dramatic drop in merger fraction from the rest-optical, where $\sim$60\% of galaxies are classified as mergers, compared to the rest-NIR, where few ($<$10\%) galaxies appear to be mergers. This wavelength dependence could in part be driven by dust attenuation, which likely leads to clumpier morphologies and could increase M20 values in shorter wavelength bands, biasing rest-optical merger fractions. Alternatively, the underlying stellar populations traced by the rest-NIR may be intrinsically smoother than the younger stellar populations observed in the rest-optical \citet{constantin2025}. While using longer wavelength bands to classify mergers is appealing due to minimal dust attenuation, PSF-smoothing effects may lead spuriously low merger fractions in the rest-NIR \citep{wang2024}.  Additionally, the \citet{Lotz2008} merger classification scheme was calibrated in the rest-optical at lower redshift \citep[\emph{z}$<$1.2, for efforts to extend this relation to $1<z<2$ in cluster galaxies see][]{sazonova2020}; disagreement in merger fraction across 1.5-4.4$\mu$m may reflect that a different relation is necessary at longer wavelengths, either due to dust effects or intrinsically different morphologies in the rest-NIR. Although no 1.1mm-selected galaxies are classified as mergers at $\lambda_{obs}>3.56\mu$m using Gini-M20, a visual inspection of the sample in longer wavelength bands reveals features such as tidal tails and multiple cores in a significant fraction ($>20\%$), indicating that several of these systems could plausibly be classified as mergers. Due to these challenges, which could lead to an overestimate in merger fraction at short wavelengths and an underestimate at long wavelengths, we strongly caution against a direct comparison of merger fractions using Gini-M20 at different wavelengths.

\begin{figure*}
    \centering
    \includegraphics[width=18cm, keepaspectratio]{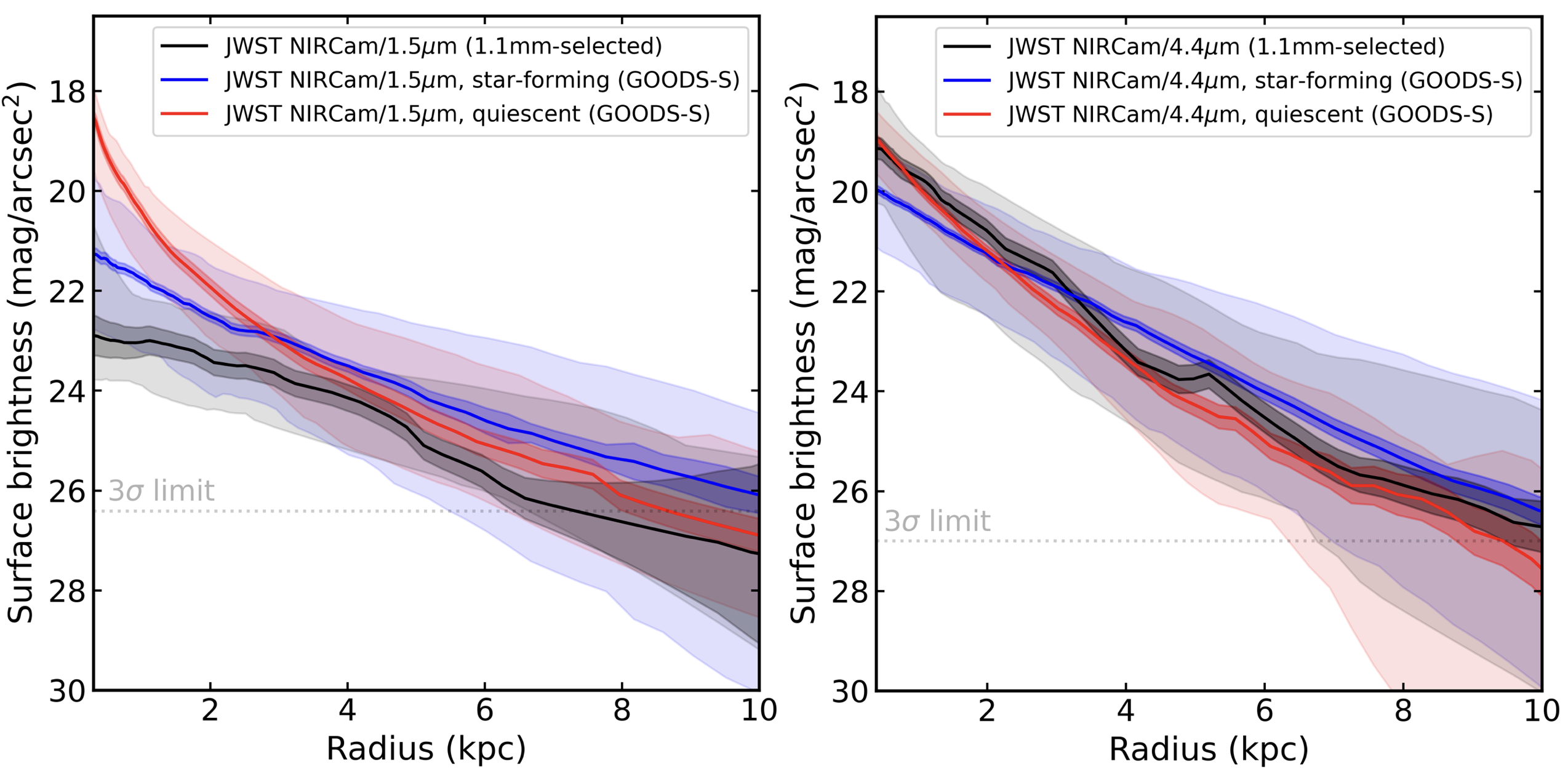}
    \caption{Left: The mean 1.5$\mu$m surface brightness profiles of quiescent (red) galaxies are brighter near the center relative to star-forming (blue), and 1.1mm-selected galaxies (black). The darker shaded area is the the standard error of the mean, and the lighter shaded area is the scatter. Right: The mean 4.4$\mu$m surface brightness profiles near the center of quiescent galaxies is very similar to that of the 1.1mm-selected sample -- both are brighter than star-forming galaxies within the central $\sim$ 3 kpc. }
    \label{fig:sbp}
\end{figure*}

The similarity between the merger fractions of the 1.1mm-selected sample and the field sample at all wavelengths indicates that although mergers may be common in both the 1.1mm-selected and massive star-forming sample, increased merger activity alone likely does not explain the enhanced star formation in the 1.1mm-selected sample. This result is consistent with \citet{gillman2024}, who find comparable merger fractions in a sample of 80 SMGs relative to more typical star-forming galaxies. \citet{gillman2024} instead suggest that disk instability from a weaker bulge component drives the growth of star formation and dust in the SMGs, since they find that the S\'ersic indices of SMGs are lower relative to more typical star-forming galaxies. However, the average S\'ersic indices of both our field ($n_{4.4\mu m}=1.49$) and 1.1mm-selected populations ($n_{4.4\mu m}$=1.44) are consistent with one another; after performing Kologonov-Smirov test, we determine that the S\'ersic indices of both populations come from the same underlying distribution.  Thus, within our sample we consider increased disk instability to be a less plausible explanation for the heightened star formation activity of the 1.1mm-selected sample. Since we do not find evidence for increased major merger activity, we favor a scenario in which minor mergers and inflows from a gas-rich environment may trigger star formation in the 1.1mm-selected sample, which is consistent with the predictions of hydrodyanmic simulations \citep{narayanan2015, lovell2021}. In turn, this star formation produces large dust reservoirs which attenuate rest-optical light. 

Whereas our analysis to this point has focused on the individual parameterized light profiles, finding no significant differences between the 1.1mm-selected and other star-forming galaxies on average in terms of S\'ersic index or merger fractions, we see qualitatively that 1.1mm-selected galaxies frequently have a compact, rest-NIR bright core. In the following section, we consider the radial light profiles at rest-optical and rest-NIR wavelengths in order to investigate the evolutionary status of the 1.1mm-selected sample.

\subsection{The onset of bulge formation in the 1.1mm-selected sample}

\label{sec:sbp}

Dusty star-forming galaxies are thought to bridge the massive quiescent and star-forming populations. We present evidence in Section \ref{sec:sersic} that the sizes of the 1.1mm-selected sample are on average typical relative to other star-forming galaxies in the rest-optical, but in the rest-NIR instead follow a similar trend to quiescent galaxies. Furthermore, we show in Section \ref{sec:morph} (Figure \ref{fig:rgb}) that the rest-NIR emission of the sample qualitatively appears to be centrally concentrated in RGB images. While these results show some morphological similarities between 1.1mm-selected and quiescent galaxies in the rest-NIR, the evolutionary status of DSFGs remains unclear.

One way to explore the distribution of stellar populations in more detail beyond the simplified metric of effective radius is to compare the rest-optical (1.5$\mu$m) and rest-NIR (4.4$\mu$m) surface brightness profiles of different galaxy populations \citep[e.g.,][]{benton2024}. In Fig~\ref{fig:sbp}, we measure surface brightness profiles of 1.1mm-selected and field galaxy samples in order to assess how their radial stellar profiles would need to evolve if an evolutionary link were to exist between dusty star-forming, star-forming, and quiescent galaxies. 
The surface brightness profiles are measured according to the method originally developed by \citet{Szomoru2010}, where the flux at a given radius is calculated as the sum of the S\'ersic flux at that radius and the flux of the residual within an aperture. We use elliptical annuli with the axis ratio and position angle determined by the best fit S\'ersic profile. 

In the rest-optical, there are noticeable differences within the central $\sim$3 kpc between the three populations. Within this region, the average profile of the 1.1mm-detected sample is notably fainter than that of the star-forming and quiescent galaxies. Given that the 1.1mm-detected sample is selected to have strong dust continuum that appears centrally concentrated, the reduced central rest-optical brightness is likely due to dust obscuration. Star-forming galaxies are also fainter at the center relative to quiescent galaxies \citep[see also][]{benton2024} -- this may either be caused by stronger dust attenuation in the central regions of star-forming galaxies relative to the quiescent population, or the difference may be intrinsic to the populations, with quiescent galaxies having more compact morphologies as a result of the quenching process \citep[e.g.,][]{wellons2015}. 

If we instead consider the rest-NIR, the surface brightness profiles of the 1.1mm-detected sample within the central $\sim$2 kpc are closely aligned with those of quiescent galaxies, whereas typical star-forming galaxies are fainter by $\sim$1 magnitude. Because the rest-NIR more robustly traces stellar mass than the rest-optical \citep[e.g.,][]{clausen2025}, this similarity suggests that many dusty star-forming galaxies have already undergone the process of bulge formation, or at least are in the process of bulge formation.  In contrast, it appears that more typical star-forming galaxies have yet to build up centrally concentrated stellar mass.  This difference is not immediately evident from the single S\'ersic profile fits, highlighting the value of examining the full radial light profile. Our analysis is similar to that presented in \citet{benton2024} with one key difference in that we explicitly divide the star-forming populationthey into two sub-categories, separating out the dusty (1mm-detected) star-forming galaxies. While we reach similar conclusion that bulge formation has taken place at this epoch, we find that the dusty star-forming galaxies within the \citet{benton2024} analysis are likely driving the result. 

We note that the surface brightness profiles
of star-forming, quiescent, and 1.1mm–selected
galaxies in Figure \ref{fig:sbp} exhibit substantial scatter within each population. However, despite this intrinsic diversity, the average profiles show systematic differences, particularly in the central regions. In the rest-NIR (F444W), 1.1mm–selected and quiescent galaxies remain more centrally concentrated than star-forming galaxies on average, indicating that these populations are not simply drawn from the same underlying distribution. That said, the 1.1mm–selected sample may be heterogeneous, and this interpretation may apply only to a subset of the population. Given the small number of galaxies in 1.1mm-selected sample, it is difficult to discern whether some subset of the 1.1mm-selected is responsible for the similarities with the quiescent population. Nonetheless, we attempt to test this by dividing the 1.1mm-selected sample along the median stellar mass (logM$_{\mathrm{star}}/$M$_\odot$=10.8).  We find tentative evidence that steeper surface brightness profiles are primarily associated with the intermediate-mass (logM$_{\mathrm{star}}/$M$_\odot<$10.8) 1.1mm-selected galaxies, while the more massive systems exhibit flatter profiles resembling the field star-forming galaxy sample.  If confirmed, this diversity could suggest that not all dusty star-forming galaxies follow a single evolutionary pathway toward massive quiescent systems.  In particular, the presence of compact rest-NIR morphologies among intermediate-mass 1.1mm–selected galaxies challenges the commonly invoked picture in which the sub-mm phase is expected to precede the formation of the most massive quiescent galaxies \citep{blain2002, nelson2014}.

Our analysis provides empirical evidence supporting the idea that massive galaxies build stellar mass from the inside out \citep[e.g.,][]{nelson2016}, and naturally fits into the framework where dusty star-forming galaxies, after undergoing an intense burst of star formation (either through secular processes or merger activity as discussed in Section \ref{sec:mergers}), are the progenitors of massive quiescent galaxies.  The quenching process therefore would not require morphological evolution; instead, once new star formation halts, dust will no longer be created and the cycle of dust destruction and depletion begins \citep[e.g.,][] {whitaker2021,lorenzon2025}. This result is also consistent with dynamical studies, which find that the $M-\sigma$ relation is similar between DSFGs and massive quiescent galaxies at low redshift \citep{birkin2021,amvrosiadis2025}. 

Alternatively, other recent studies of dusty star-forming galaxies find evidence that this population is at the onset of bulge formation, with stellar mass assembly at the core still ongoing \citep[e.g.][]{chen2022,hodge2025}. In a sample of 22 dusty star-forming galaxies, \citet{lebail2024} find that 7 had a quiescent bulge and a star forming disk, while 10 had a star-forming core and disk, and 5 had a star-forming core embedded in a quiescent disk. Our 1.1mm-selected sample has a higher average stellar mass that is more consistent with the quiescent bulge population, and may be dominated by this subset of dusty star-forming galaxies. Given the morphological diversity seen in larger samples of dusty star-forming galaxies \citep[e.g.][]{gillman2024,ren2025}, it is likely that this population represents a range of evolutionary stages, as suggested by the difference in rest-NIR surface profiles with stellar mass noted earlier. A caveat to our analysis of the rest-NIR surface brightness profiles is that in dusty star-forming systems, the rest-NIR may be tracing dust-obscured star formation in addition to the bulk stellar population \citep{nelson2016b, miller2022}. In this case, there remain residual outshining effects compounded with extreme dust attenuation that could flatten surface brightness profiles in the most massive and dusty galaxies. We therefore cannot rule out ongoing bulge formation rather than the interpretation that the bulge is already in place. Nevertheless, our sample adds to the growing evidence that the DSFG phase may be a key turning point in stellar mass assembly, in which galaxies have completed or are on the cusp of bulge formation. 

Spatially resolved studies of the molecular gas in these systems could potentially further clarify the evolutionary status of these systems. \citet{park2022} predict that there is a relationship between the quenching timescale with the distribution of molecular gas relative to stellar mass; additionally, galaxies undergoing rapid quenching may transform morphologically to a compact elliptical prior to or after quenching depending on whether a major merger occurs. Studying the extent and kinematics of molecular gas could help determine where and when these massive galaxies quench.

\section{Conclusion}

\label{sec:conclusion}

In a multi-wavelength study of 33 1.1mm-selected galaxies at $1.5<z<5.5$ in GOODS-S, we use JWST/NIRCam and ALMA imaging to study the morphological properties of these galaxies in comparison to field samples of similarly massive star-forming and quiescent galaxies at $1<z<5.5$.

We find that:
\begin{itemize}
    \item The 1.1mm-selected galaxies are mostly similar in the rest-optical and compact in the rest-NIR relative to the sizes of typical star-forming galaxies: $\approx$30\% fall above the average size-mass relation at 4.4$\mu$m. There is a 2.6$\times$ steeper size-wavelength gradient for the 1.1mm-selected sample relative to the overall massive star-forming galaxy population.
    \item Dust (as traced by 1.1mm) is frequently co-spatial with the bulk stellar population (as traced by 4.4$\mu$m) and is located in compact central regions. The steep size-wavelength gradient in 1.1mm-selected galaxies is a direct result of dust obscuration in the cores of these galaxies.
    \item We find that the rest-NIR light is more extended relative to dust emission at 1.1mm.  This result confirms other recent studies using 870$\mu$m dust sizes, but now tracing longer mm wavelengths safely in the Raleigh-Jeans regime at $>$250$\mu$m up to $z=3.4$ (versus $z=2.5$ previously).
    \item Using Gini-M20 parameters, we measure similar merger fractions between the 1.1mm-selected sample and other star-forming galaxies, but significant variation ($\sim$0\% to 60\%) in merger fraction as a function of wavelength from the rest-NIR to the rest-optical. We also do not find clear evidence for enhanced disk instability in the 1.1mm-selected sample. Therefore, we favor cold accretion and minor mergers as responsible for enhanced star formation in the 1.1mm-selected sample. 
    \item The rest-NIR surface brightness profiles within the central $\sim$3 kpc are strikingly similar for quiescent and 1.1mm-selected galaxies. This suggests that galaxies within the 1.1mm-selected sample may have developed a central bulge and supports an inside-out model of galaxy growth where these dusty star-forming galaxies are the direct progenitors of early massive quiescent galaxies.
\end{itemize}

In summary, our results indicate that the 1.1mm-selected galaxies have distinct rest-optical and rest-NIR morphologies when compared to more typical massive star-forming galaxies. Morphological differences between these two populations are driven at least in part by dust attenuation in central, compact star-forming regions. The stellar mass distributions of the 1.1mm-selected sample, indicate that unlike their more typical star-forming counterparts, stellar mass assembly may be nearly complete, as they have already formed a central bulge. While we speculate that this subset of massive galaxies may be on the precipice of quenching, future spectroscopic analyses that explicitly compare the kinematics of these two populations will be critical to validate this scenario.  Moreover, resolved studies probing the molecular gas reservoirs of these dusty systems will be invaluable for placing strict limits on the the star formation efficiency and possible quenching timescales should no further cold gas be accreted. 

\section{Acknowledgements}

We thank the anonymous referee for helpful comments which have improved the quality of this work. The University of Massachusetts Amherst acknowledges that it was founded and built on the unceded homelands of the Pocumtuc Nation on the land of the Norrwutuck community. This work is based in part on observations made with the NASA/ESA/CSA James Webb Space Telescope and the NASA/ESA Hubble Space Telescope obtained from the Space Telescope Science Institute, which is operated by the Association of Universities for Research in Astronomy, Inc., under NASA contract NAS 5–26555. The data were obtained from the Mikulski Archive for Space Telescopes at the Space Telescope Science Institute, which is operated by the Association of Universities for Research in Astronomy, Inc., under NASA contract NAS 5-03127 for JWST. These observations are associated with programs JWST-GO-1895, JWST-GO-1963, and JWST-GTO-1210. KEW and SB gratefully acknowledge support for program JWST-GO-1895 provided by NASA through a grant from the Space Telescope Science Institute, which is operated by the Associations of Universities for Research in Astronomy, Incorporated, under NASA contract NAS 5-03127, as well as support from NSF-CAREER\#2144314.  I.S. acknowledges fundings from the European Research Council (ERC) DistantDust (Grant No.101117541) and the Atraccíon de Talento Grant No.2022-T1/TIC-20472 of the Comunidad de Madrid, Spain. Some of the data products presented herein were retrieved from the Dawn JWST Archive (DJA). DJA is an initiative of the Cosmic Dawn Center (DAWN), which is funded by the Danish National Research Foundation under grant DNRF140.
ALMA data were retrieved from ECOGAL (ECOlogy for Galaxies using ALMA archive and Legacy surveys) program (Marie Skłodowska-Curie grant agreement No 101107795).
This work has received funding from the Swiss State Secretariat for Education, Research and Innovation (SERI) under contract number MB22.00072, as well as from the Swiss National Science Foundation (SNSF) through project grant 200020\_207349. 
\bibliography{bib}{}
\bibliographystyle{aasjournal}

\end{document}